\documentclass[twocolumn]{aastex631}

\usepackage{hyperref,enumerate}
\usepackage{xcolor}

\hypersetup{
    unicode=false,                  
    pdftoolbar=true,                
    pdfmenubar=true,                
    pdffitwindow=true,              
    pdfstartview={FitH},            
    pdftitle={BD+44 493 NEID},      
    pdfauthor={vplacco},            
    pdfsubject={Astronomy},         
    pdfcreator={dvipdf},            
    pdfproducer={dvipdf},           
    pdfkeywords={metal-poor stars}, 
    pdfnewwindow=true,              
    colorlinks=true,                
    linkcolor=red,                  
    citecolor=blue,                 
    filecolor=magenta,              
    urlcolor=cyan,                  
    breaklinks=true,
    linktocpage
}

\definecolor{emerald}{rgb}{0.31, 0.78, 0.41}

\newcommand{\kmsec}{\mbox{km~s$^{\rm -1}$}}
\newcommand{\msec}{\mbox{m~s$^{\rm -1}$}}
\newcommand{\eps}[1]{\ensuremath{\log\epsilon\,(\mathrm{#1})}}
\newcommand{\vv}{{\tablenotemark{\footnotesize{a}}}}

\newcommand{\abund}[2]{\ensuremath{[\mathrm{#1}/\mathrm{#2}]}}
\newcommand{\cfe}{\abund{C}{Fe}}

\newcommand{\xfe}[1]{\abund{#1}{Fe}}

\newcommand{\metal}{\abund{Fe}{H}}

\newcommand{\teff}{\ensuremath{T_\mathrm{eff}}}
\newcommand{\logg}{\ensuremath{\log\,g}}
\newcommand{\Msun}{M$_\odot$}

\newcommand{\A}[1]{\ensuremath{A(\mathrm{#1})}}

\newcommand{\K}{\ensuremath{\mathrm{K}}}

\newcommand{\ravel}{\object{BD$+$44~493}}
\newcommand{\rave}{\object{\mbox{BD$+$44$^{\circ}$493}}}

\defcitealias{sestito2019}{S+19}

\received{September 6, 2024}
\revised{October 9, 2024}
\accepted{October 10,2024}

\submitjournal{ApJ}

\begin{document}

\title{\rave: Chemo-Dynamical Analysis and Constraints on \\ Companion Planetary Masses from WIYN/NEID Spectroscopy\footnote{The WIYN Observatory is a joint facility of the University of Wisconsin–Madison, Indiana University, NSF NOIRLab, the Pennsylvania State University, Purdue University, and Princeton University.
Based on data collected at the Subaru Telescope, which is operated by the
National Astronomical Observatory of Japan.
Based on observations made with the NASA/ESA Hubble Space Telescope,
obtained at the Space Telescope Science Institute, which is operated by the
Association of Universities for Research in Astronomy, Inc., under NASA
contract NAS 5-26555. These observations are associated with programs
GO-12268, GO-12554, and GO-14231.
}}


\author[0000-0003-4479-1265]{Vinicius M.\ Placco}
\affiliation{NSF NOIRLab, Tucson, AZ 85719, USA}

\author[0000-0002-5463-9980]{Arvind  F.\ Gupta}
\affiliation{NSF NOIRLab, Tucson, AZ 85719, USA}

\author[0000-0002-8048-8717]{Felipe Almeida-Fernandes}
\affiliation{Departamento de Astronomia, Instituto de Astronomia, Geof\'isica e
Ci\^encias Atmosf\'ericas da USP, Cidade \\ Universit\'aria, 05508-900, S\~ao
Paulo, SP, Brazil}
\affiliation{Observat\'orio do Valongo, Ladeira Pedro Ant\^onio, 43, Sa\'ude, Rio de Janeiro, 20080-090, BR}

\author[0000-0002-9632-9382]{Sarah E.\ Logsdon}
\affiliation{NSF NOIRLab, Tucson, AZ 85719, USA}

\author[0000-0002-2488-7123]{Jayadev Rajagopal}
\affiliation{NSF NOIRLab, Tucson, AZ 85719, USA}

\author[0000-0002-5463-6800]{Erika M.\ Holmbeck}
\affiliation{Lawrence Livermore National Laboratory, Livermore, CA 94550, USA}
\affiliation{Joint Institute for Nuclear Astrophysics -- Center for the Evolution of the Elements (JINA-CEE), USA}

\author[0000-0001-5107-8930]{Ian U.\ Roederer}
\affiliation{Department of Physics, North Carolina State University, Raleigh, NC 27695, USA}
\affiliation{Joint Institute for Nuclear Astrophysics -- Center for the Evolution of the Elements (JINA-CEE), USA}


\author[0000-0003-0928-2000]{John Della Costa}
\affiliation{NSF NOIRLab, Tucson, AZ 85719, USA}

\author{Pipa Fernandez}
\affiliation{NSF NOIRLab, Tucson, AZ 85719, USA}

\author[0009-0003-8107-4764]{Eli Golub}
\affiliation{NSF NOIRLab, Tucson, AZ 85719, USA}

\author[0000-0002-3985-8528]{Jesus Higuera}
\affiliation{NSF NOIRLab, Tucson, AZ 85719, USA}

\author[0009-0004-7817-2547]{Yatrik Patel}
\affiliation{NSF NOIRLab, Tucson, AZ 85719, USA}

\author{Susan Ridgway}
\affiliation{NSF NOIRLab, Tucson, AZ 85719, USA}

\author{Heidi Schweiker}
\affiliation{NSF NOIRLab, Tucson, AZ 85719, USA}


\correspondingauthor{Vinicius M.\ Placco}
\email{vinicius.placco@noirlab.edu}

\begin{abstract}

In this work, we present high-resolution ($R\sim100,000$), high signal-to-noise ($\rm{S/N}\sim800$) spectroscopic observations for the well-known, bright, extremely metal-poor, carbon-enhanced star \rave. We determined chemical abundances and upper limits for 17 elements from  WIYN/NEID data, complemented with 11 abundances re-determined from Subaru and Hubble data, using the new, more accurate, stellar atmospheric parameters calculated in this work. 
Our analysis suggests that \rave\, is a low-mass ($0.83~\rm{M}_{\odot}$) old ($12.1-13.2$~Gyr) second-generation star likely formed from a gas cloud enriched by a single metal-free 20.5\,M$_\odot$ Population III star in the early Universe. 
With a disk-like orbit, \rave\, does not appear to be associated with any major merger event in the early history of the Milky Way.
From the precision radial-velocity NEID measurements (median absolute deviation -- MAD~=~16~\msec), we were able to constrain companion planetary masses around \rave\, and rule out the presence of planets as small as $m \sin i = 2$M$_{\rm J}$ out to periods of 100 days. This study opens a new avenue of exploration for the intersection between stellar archaeology and exoplanet science using NEID.
\
\end{abstract}


\section{Introduction} 
\label{intro}

Extremely and Ultra Metal-Poor \citep[EMP and UMP -- \metal\footnote{\abund{A}{B} = $\log(N_A/{}N_B)_{\star} - \log(N_A/{}N_B) _{\odot}$, where $N$ is the number density of atoms of a given element in the star ($\star$) and the Sun ($\odot$), respectively.}$\leq-3.0$ and $\leq-4.0$, respectively;][]{frebel2015} stars are ``local'', low-mass objects with cosmological significance, carrying in their atmospheres the chemical byproducts of the explosive evolution of the very first generation of stars (Population III - Pop. III) to be formed in the early universe \citep{abel2002,bromm2004,bromm2009}. 
Through the study of the chemical abundance patterns of these bonafide second-generation stars, it is possible to infer the main characteristics of the \emph{first} initial mass function, which in turn can constrain the existence of surviving low-mass Pop. III stars in the Milky Way and its satellite galaxies \citep{klessen2023,koutsouridou2024}.

These ``rare gems'' are hard to find: observational evidence suggests that only 1 in 10,000 stars in the solar neighborhood are EMP, and there is roughly one UMP star with $V<18$~mag for every 100~deg$^2$ of the Galactic halo \citep{youakim2017}. In addition, due to the decreasing strength of the iron absorption features with decreasing metallicity, stars with \metal$\leq-5.0$ only have a handful of lines available for measurement in the optical wavelength regime, requiring high-resolution, high signal-to-noise ratio (S/N) observations\footnote{As an example, HE~1327$-$2326, an unevolved halo star with \metal=$-5.45$, had only seven \ion{Fe}{1} lines detected in the $3500\leq$$\lambda$(\AA)$\leq4000$ range, from high-resolution ($R\sim60,000$), high S/N ($\sim$100 at 4000\AA) data, with equivalent widths between 1.9 and 6.8 m\AA\, \citep{aoki2006}.}.
An alternative approach to search for low-metallicity stars is to find objects presenting enhancements in carbon \citep{rossi2005,placco2010,placco2011}. It has been recognized that the fraction of Carbon-Enhanced Metal-Poor \citep[CEMP - \metal$<-1.0$ and \cfe$>+0.7$;][]{aoki2007} stars increases for decreasing metallicities, from 20\% for \metal$\leq-2.0$ to 81\% for \metal$\leq-4.0$ \citep{lucatello2005,frebel2006,lee2013,placco2014c,yoon2018,arentsen2022}. A few notable exceptions include: SDSS~J102915$+$172927 \citep{caffau2011b}, CD$-$38~245 \citep{bessell1984,cayrel2004}, SDSS~J131326$-$001941 \citep{frebel2015b}, SPLUS~J210428$-$004934 \citep{placco2021b}, and AS0039 \citep{skuladottir2021}; all with \metal$\leq-4.0$ and \cfe$\lesssim0.0$.

One aspect that has not yet been fully explored in the study of EMP, UMP, and CEMP stars is the presence of exoplanet companions.
Giant planet occurrence rates are relatively low for stars with sub-solar metallicity \citep{Santos2004,Fischer2005,Boley2021}.
This metallicity-occurence rate correlation is largely informed by studies of short-period giant planets, whose larger known populaton carries greater statistical weight. Whether the trend persists for longer-period, cold Jupiters is unclear; some results indicate that, on average, cold Jupiters host stars are more metal-poor than the hosts of their short-period counterparts \citep{Petigura2018,Banerjee2024}, while other recent work focusing on cold Jupiters hosts with super-Earth companions finds these stars to be metal-rich \citep{Zhu2024}. But in either case, the most metal-poor stars known to host exoplanets have \metal$\gtrsim-0.6$ \citep{Hellier2014,Johnson2018,Polanski2021,Brinkman2023,Dai2023}.
This lower bound is consistent with theoretical expectations for the dependence of planet formation on metallicity \citep{Ida2004,Mordasini2012,Andama2024}; planet formation via core accretion is inhibited by shorter disk lifetimes \citep{Ercolano2010,Yasui2010} and lower heavy element abundances in metal-poor environments \citep{Johnson2012}. However, planet formation via disk fragmentation and collapse does not exhibit this same metallicity dependence \citep[e.g.,][]{Boss2002}, and thus the detection of planets around EMP stars may be key evidence for this formation mechanism. But exoplanet occurrence in the EMP regime has yet to be explored empirically. 


Studies of metallicity dependence for the transiting planet population have largely been restricted to \metal$\geq-1$ \citep{Guo2017,Petigura2018,Kutra2021,Boley2021,Zink2023,Boley2024}, and radial velocity and direct imaging surveys are limited to bright, nearby stars, very few of which qualify as EMP. Radial velocity studies are further limited by other selection effects as well. The measurement precision of radial velocity observations scales with the information content intrinsic to the stellar spectrum \citep{Bouchy2001}, so the dearth of lines for metal-poor stars raises the achievable precision floor.



\rave\footnote{\href{https://simbad.u-strasbg.fr/simbad/sim-id?Ident=BD\%2B44+493}{https://simbad.u-strasbg.fr/simbad/sim-id?Ident=BD\%2B44+493}} is the brightest ($V=9.075$~mag) star with $\metal\sim-4.0$ observed to date. It has been extensively studied in the literature (104 references listed on the SIMBAD database) and it was first classified as an OB$^+$ star by \citet{balazs1965}. Later on, \rave\, was re-discovered by \citet{bidelman1985} as an ``extremely weak-lined'' star from moderate-dispersion objective-prism plates taken with the Burrell Schmidt Telescope.
The first metallicity determination was published by \citet{carney1986} (\metal=$-2.9$), using the photometric $(b-y)$ metallicity method by \citet{bond1980}. \citet{anthony1994} refined the photometric estimate to \metal=$-2.71$, recognizing \rave\, as a nearby ($D=406$~pc) halo red giant.
A decade later, \citet{carney2003}\footnote{Carney and colleagues labeled \rave\, as a ``star with special problems''} used the derived stellar parameters at the time (\teff=5510~K, \logg=2.6, and \metal=$-2.71$) to select an optimal synthetic spectrum as a template to calculate radial velocities. These parameters implied that \rave\, was a Red Horizontal Branch (RHB) star. From the template fitting, Carney et al. noticed that a better correlation was produced when using a warmer (\teff=6000~K) and more metal-poor (\metal=$-3.0$) synthetic spectrum. \citet{ito2009} were the first to recognize \rave\, as an extremely metal-poor star with \metal=$-3.7$. The authors attribute the presence of strong carbon CH bands as the culprit for the photometric metallicity estimates producing values 10 times larger than the spectroscopic ones.

Due to its brightness, \rave,\ became a template for studying stars in the \metal$\sim-4.0$ regime. \citet{ito2013} expanded the work started by \citet{ito2009} using optical high-resolution, high signal-to-noise ratio (S/N) spectra obtained with the High Dispersion Spectrograph (HDS) at the Subaru Telescope. In parallel, \citet{takeda2013} obtained near-infrared data, using the Infrared Camera and Spectrograph (IRCS) at Subaru, for a sample of metal-poor stars (including \rave) and determined carbon abundances from \ion{C}{1} absorption features. \citet{aoki2015} also reported on the abundances of carbon and oxygen for \rave, calculated from equivalent width analysis of CH (3870-4356\AA) and OH (3081-3264\AA) features. In the near ultra-violet regime, Hubble Space Telescope (HST) data was used by \citet{placco2014b} (Space Telescope Imaging Spectrograph -- STIS) and \citet{roederer2016} (Cosmic Origins Spectrograph -- COS) to supplement the chemical inventory of \rave, including the first detection of phosphorus and sulfur.
The studies mentioned above agree that \rave\, is a prime candidate to be a bonafide second-generation star, likely formed from a gas cloud enriched by the explosion of a single supernova in the early universe. 

This article presents a chemo-dynamical analysis and constraints on companion planetary masses for \rave\, from high-resolution, high signal-to-noise optical spectra acquired with the NEID spectrograph. These data, coupled with 6D parameters from Gaia DR3 \citep{gaia23dr3}, allowed for the determination of more accurate stellar atmospheric parameters, chemical abundances, mass, and age. The precise radial velocities measured are used to limit, for the first time, the masses of a possible planetary companion around a \metal$\sim-4$ star.

This work is outlined as follows: Section~\ref{observations} describes the observations and processing of the NEID data, followed by details on the radial velocity measurements in Section~\ref{radvel}. The determination of stellar atmospheric parameters, chemical abundances, and a comparison with the work of \citet{ito2013} are shown in Section~\ref{atmpars}. In Sections~\ref{chemod} and \ref{planet} we discuss the chemo-dynamical nature and present constraints on planetary masses around \rave, respectively. Conclusions and perspectives for future work are given in Section~\ref{conclusion}.

\begin{deluxetable*}{lllll}
\tablecaption{Properties of \protect\ravel \label{starinfo}}
\tablewidth{0pt}
\renewcommand{\arraystretch}{1.0}
\tabletypesize{\scriptsize}
\tabletypesize{\small}
\tablehead{
\colhead{Quantity} &
\colhead{Symbol} &
\colhead{Value} &
\colhead{Units} &
\colhead{Reference}}
\startdata
Right ascension            & $\alpha$ (J2000)    & 02:26:49.74               & hh:mm:ss.ss       & \citet{gaia23dr3}             \\
Declination                & $\delta$ (J2000)    & $+$44:57:46.5             & dd:mm:ss.s        & \citet{gaia23dr3}             \\
Galactic longitude         & $\ell$              & 140.134                   & degrees           & \citet{gaia23dr3}             \\
Galactic latitude          & $b$                 & $-$14.699                 & degrees           & \citet{gaia23dr3}             \\
Gaia DR3 ID                &                     & 341511064663637376        &                   & \citet{gaia23dr3}             \\
Parallax                   & $\varpi$            & 4.8661 $\pm$ 0.0226       & mas               & \citet{gaia23dr3}             \\
Inverse parallax distance  & $1/\varpi$          & 206.8$^{+1.0}_{-0.9}$     & pc                & This study\vv                 \\
Distance                   & $d$                 & 204.3$^{+1.1}_{-1.0}$     & pc                & \citet{Bailer-Jones+2021}     \\
Proper motion ($\alpha$)   & PMRA                & 118.221 $\pm$ 0.021       & mas yr$^{-1}$     & \citet{gaia23dr3}             \\
Proper motion ($\delta$)   & PMDec               & $-$32.068 $\pm$ 0.020     & mas yr$^{-1}$     & \citet{gaia23dr3}             \\
$V$ magnitude              & $V$                 &  9.075 $\pm$ 0.005        & mag               & \citet{henden2014}            \\
$G$ magnitude              & $G$                 &  8.863 $\pm$ 0.003        & mag               & \citet{gaia23dr3}             \\
$BP$ magnitude             & $BP$                &  9.209 $\pm$ 0.003        & mag               & \citet{gaia23dr3}             \\
$RP$ magnitude             & $RP$                &  8.327 $\pm$ 0.004        & mag               & \citet{gaia23dr3}             \\
Color excess               & $E(B-V)$            & 0.0230 $\pm$ 0.0009       & mag               & \citet{schlafly2011}          \\
Bolometric correction      & BC$_V$              & $-$0.50 $\pm$ 0.03        & mag               & \citet{casagrande2014}        \\
\hline
Signal-to-noise ratio           @3800\AA  & S/N  & 93                        & pixel$^{-1}$      & This study                    \\
\phantom{Signal to noise ratio} @4560\AA        && 432                       & pixel$^{-1}$      & This study                    \\
\phantom{Signal to noise ratio} @5180\AA        && 574                       & pixel$^{-1}$      & This study                    \\
\phantom{Signal to noise ratio} @6580\AA        && 797                       & pixel$^{-1}$      & This study                    \\
\hline
Effective Temperature      & \teff               & 5351 $\pm$ 51             & K                 & This study                    \\
                           &                     & 5430 $\pm$ 150            & K                 & \citet{ito2013}               \\
Log of surface gravity     & \logg               & 3.12 $\pm$ 0.07           & (cgs)             & This study                    \\
                           &                     & 3.40 $\pm$ 0.30           & (cgs)             & \citet{ito2013}               \\
Microturbulent velocity    & $\xi$               & 1.45 $\pm$ 0.10           & \kmsec            & This study                    \\
                           &                     & 1.30 $\pm$ 0.30           & \kmsec            & \citet{ito2013}               \\
Metallicity                & \metal              & $-$3.96 $\pm$ 0.09        & dex               & This study                    \\
                           &                     & $-$3.83 $\pm$ 0.19        & dex               & \citet{ito2013}               \\
\hline
Radial velocity            & RV                  & $-$150.445 $\pm$ 0.016    & \kmsec            & This study                    \\ 
Isochronal-based age      &                     & $10.3_{-3.5}^{+2.9}$      & Gyr               & This study                    \\
Kinematical-based age      &                     & $13.1_{-0.9}^{+0.5}$      & Gyr               & This study                    \\
Constrained age range     &                     & $[12.1,13.2]$      & Gyr               & This study                    \\
Mass                       & $M$                 & $0.83_{-0.05}^{+0.09}$    & $M_{\odot}$       & This study                    \\
\hline
Galactocentric coordinates & ($X,Y,Z$)           & ($+8.36,+0.13,-0.03$)     & kpc               & This study                    \\
Galactic space velocity    & ($U,V,W$)           & ($+33.5,-181.5,+51.5$)    & km s$^{-1}$       & This study                    \\
Total space velocity       & $V_\mathrm{Tot}$    & $+191.6$                  & km s$^{-1}$    & This study                    \\
Apogalactic radius         & $R_\mathrm{apo}$    & $+$8.623 $\pm$ 0.002      & kpc            & This study                    \\
Perigalactic radius        & $R_\mathrm{peri}$   & $+$1.366 $\pm$ 0.011      & kpc            & This study                    \\
Max. distance from the Galactic plane & $z_\mathrm{max}$ & 1.286 $\pm$ 0.001 & kpc            & This study                    \\
Orbital eccentricity       & $ecc$               & 0.726 $\pm$ 0.002         &                   & This study                    \\
Vertical angular momentum  & $L_Z$               & ($+0.728 \pm 0.004) \cdot 10^3$  & kpc km s$^{-1}$   & This study               \\
Total orbital energy       & $E$                 & ($-1.779 \pm 0.001) \cdot 10^5$  & km${^2}$ s$^{-2}$ & This study               \\
\enddata
\tablenotetext{a}{Using $\varpi_{\rm zp} = -0.0314$ mas from \citet{lindegren2020}.}                                               
\end{deluxetable*}

\begin{figure*}
 \includegraphics[width=1\linewidth]{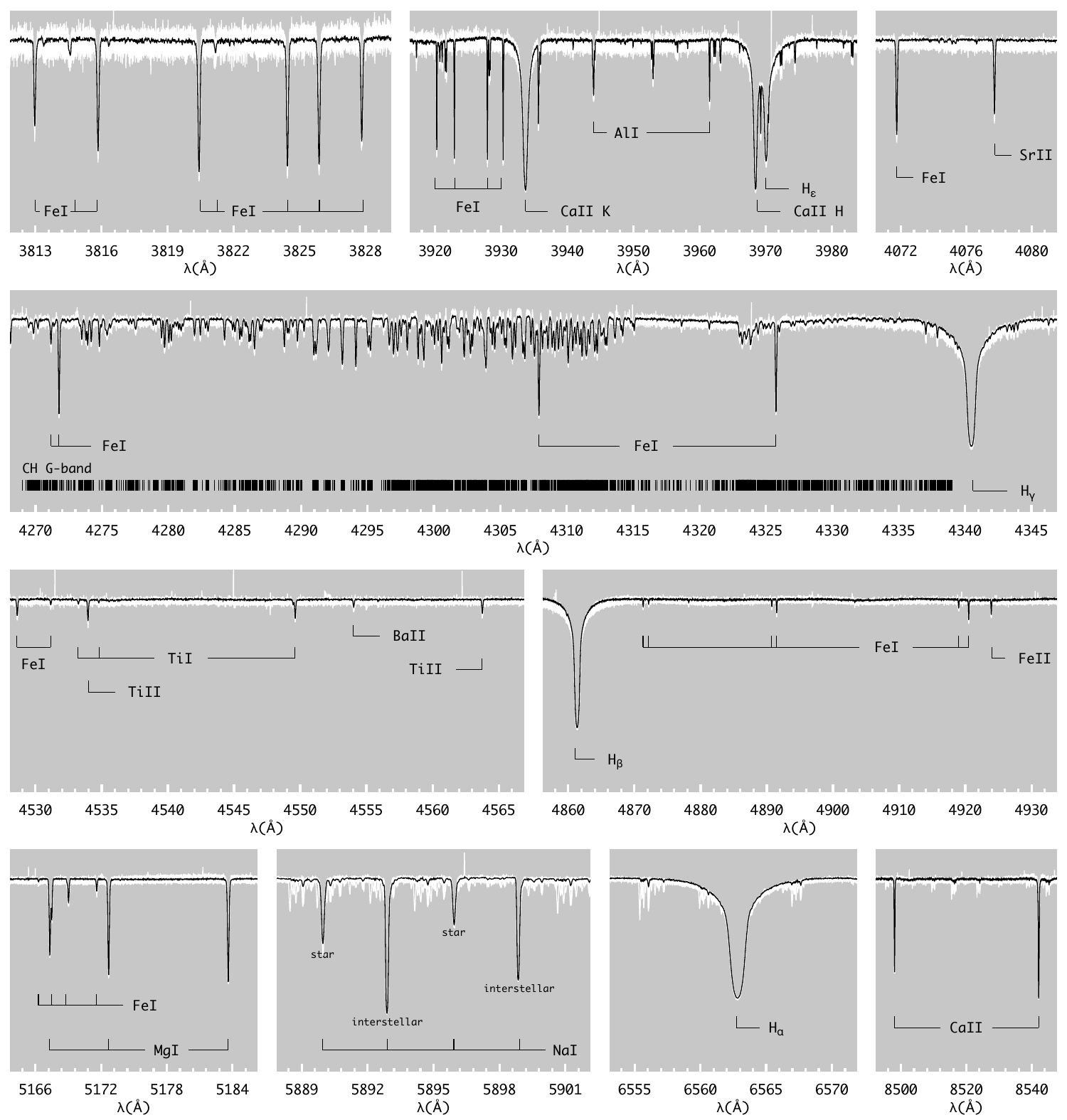}
     \caption{Normalized flux as a function of wavelength (in \AA) for selected regions of the NEID spectra, highlighting a few absorption features of interest for stellar parameter and chemical abundance calculations. The solid black line represents the combined spectra and the white lines show the range of normalized fluxes for the individual exposures. }
 \label{combo}
\end{figure*}

\section{Observations}
\label{observations}

\rave\ was observed with NEID\footnote{Prop. ID 2023B-879248; PIs: J. Rajagopal and V. Placco.}
\citep{schwab2016}, a fiber-fed \citep{kanodia2023}, environmentally-stabilized \citep{robertson2019} echelle spectrograph on the WIYN 3.5-meter Telescope at NSF Kitt Peak National Observatory. Observations were conducted in the high-resolution mode\footnote{There was one 1200 s exposure taken in the high-efficiency mode (HE - $R\sim60,000$) as a test on the first night of observations.} (HR - $R\sim110,000$), which was chosen due to its better resolution for similar radial velocity precision and S/N when compared to the high-efficiency mode. Data were taken on 14 separate nights between 2023 October 3 and 2024 January 28, and we collected a single 1800 s exposure each night. The NEID echelle spectra were processed with version $1.3$ of the NEID Data Reduction Pipeline\footnote{\url{https://neid.ipac.caltech.edu/docs/NEID-DRP/}} (DRP), producing wavelength-calibrated 1D spectra, radial velocity measurements and uncertainties, and other higher level data products. Table~\ref{starinfo} lists basic photometric and astrometric data, as well parameters for \rave\, derived in this work.

\begin{deluxetable*}{ccccccccr}[!ht] 
\tablewidth{0pc}
\tablecaption{\label{rvtab} Radial velocity information}
\tablehead{
\colhead{Date}&
\colhead{Time}&
\colhead{Exp. time} &
\colhead{Airmass}&
\colhead{Julian date}&
\colhead{Obs.\ Mode}&
\colhead{RV}&
\colhead{$\sigma_{\rm RV}$}&
\colhead{S/N}\\
\colhead{}&
\colhead{(UTC)}&
\colhead{(s)}&
\colhead{}&
\colhead{}&
\colhead{}&
\colhead{(km s$^{-1}$)}&
\colhead{(km s$^{-1}$)}&
\colhead{(pixel$^{-1}$)}}
\startdata
2024-01-28 & 04:15:27 & 1800  & 1.2201 &  2460337.69002875 & HR & $-$150.4462 & 0.0337 &  44 \\
2024-01-14 & 05:15:29 & 1800  & 1.2333 &  2460323.73178155 & HR & $-$150.4336 & 0.0232 &  69 \\
2024-01-07 & 04:07:18 & 1800  & 1.0637 &  2460316.68573550 & HR & $-$150.4455 & 0.0635 &  21 \\
2023-12-31 & 04:13:11 & 1800  & 1.0456 &  2460309.69044952 & HR & $-$150.4144 & 0.0219 &  71 \\
2023-12-22 & 04:51:54 & 1800  & 1.0480 &  2460300.71730500 & HR & $-$150.4371 & 0.0226 &  69 \\
2023-12-13 & 02:39:05 & 1800  & 1.1024 &  2460291.62633709 & HR & $-$150.4366 & 0.0218 &  72 \\
2023-12-04 & 06:08:37 & 1800  & 1.0526 &  2460282.77121219 & HR & $-$150.4678 & 0.0184 &  87 \\
2023-11-23 & 04:34:29 & 1800  & 1.0591 &  2460271.70706967 & HR & $-$150.4540 & 0.0180 &  86 \\
2023-11-01 & 07:55:54 & 1800  & 1.0377 &  2460249.84659339 & HR & $-$150.4198 & 0.0240 &  64 \\
2023-10-30 & 09:46:34 & 1800  & 1.1665 &  2460247.92237922 & HR & $-$150.4202 & 0.0230 &  68 \\
2023-10-29 & 10:43:40 & 1800  & 1.3114 &  2460246.96329049 & HR & $-$150.4442 & 0.0193 &  83 \\
2023-10-28 & 07:05:52 & 1800  & 1.0300 &  2460245.81155101 & HR & $-$150.4562 & 0.0183 &  86 \\
2023-10-26 & 09:45:30 & 1800  & 1.1338 &  2460243.92258020 & HR & $-$150.4452 & 0.0255 &  63 \\
2023-10-03 & 10:24:52 & 1200  & 1.0637 &  2460220.94544959 & HE & $-$150.5173 & 0.0190 & 145 \\
2023-10-03 & 09:51:50 & 1800  & 1.0386 &  2460220.92647457 & HR & $-$150.4725 & 0.0153 & 157 \\
\enddata
\end{deluxetable*}

\begin{figure*}
 \includegraphics[width=1\linewidth]{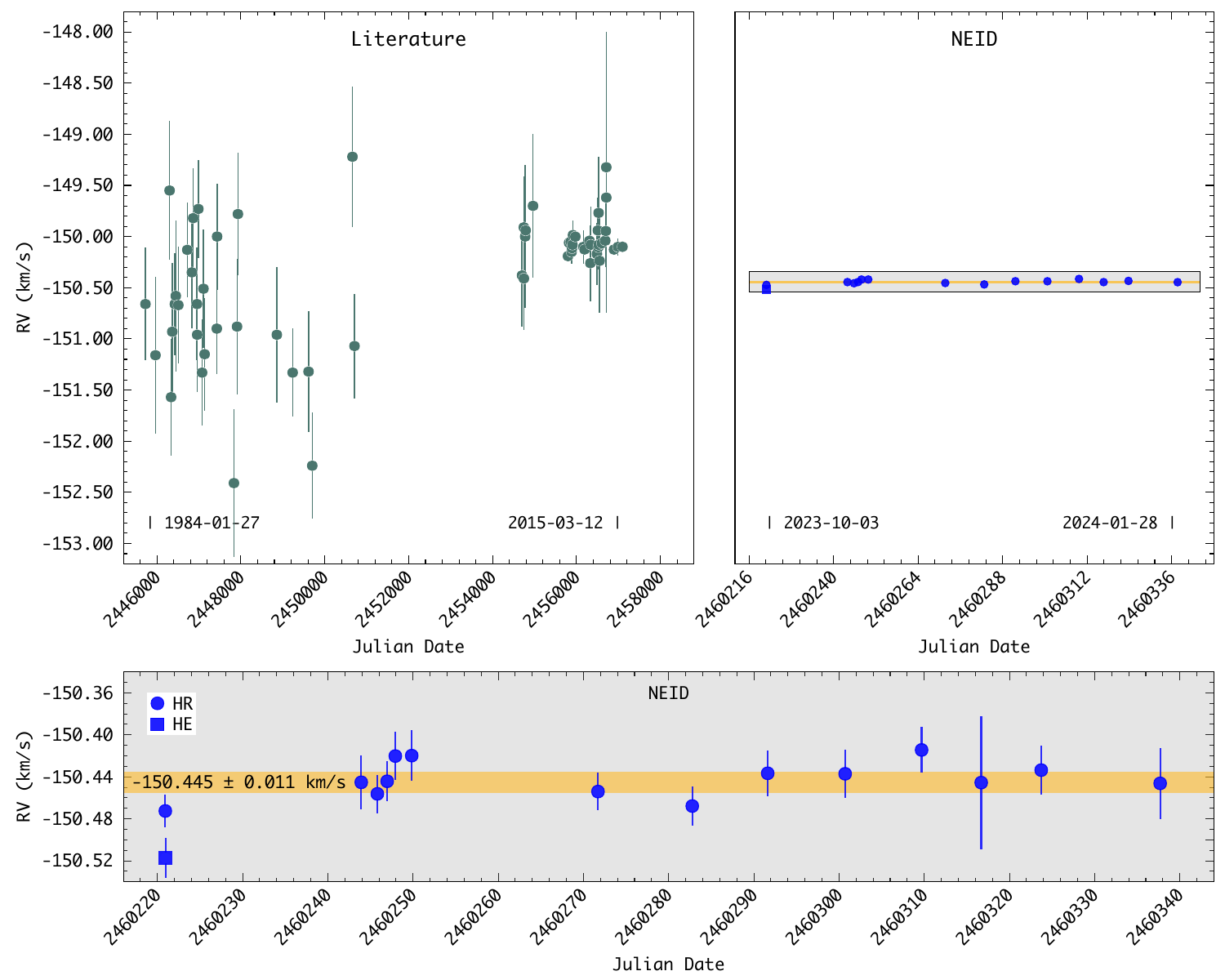}
     \caption{Top: Radial velocities for \rave\, from the literature (left) and from NEID (right). Individual measurements and references can be found in Tables~\ref{rvtab} and \ref{rvlit}. Bottom: Zoom in for the NEID radial velocities, with the median and median absolute deviation values annotated. The thickness of the horizontal line is proportional to 1$\sigma$.}
 \label{rvfig}
\end{figure*}

Each of the 14 HR exposures had their echelle orders normalized individually using the software {\texttt{smhr}}\footnote{\href{https://github.com/andycasey/smhr}{https://github.com/andycasey/smhr}} \citep{casey2014}, which produced a 1D spectrum through variance-weighting stitching of overlapping orders. The normalized, radial velocity corrected spectra were combined using the {\texttt{scombine}} task in \texttt{NOIRLab IRAF}\footnote{NOIRLab IRAF is distributed by the Community Science and Data Center at NSF NOIRLab, which is managed by the Association of Universities for Research in Astronomy (AURA) under a cooperative agreement with the U.S. National Science Foundation.} with a simple average and sigma-clipping rejection. The final signal-to-noise ratios per pixel in selected spectral regions are listed in Table~\ref{starinfo}. The panels in Figure~\ref{combo} show portions of the NEID combined data, highlighting absorption features of interest for stellar atmospheric parameter and chemical abundance work, as described in Section~\ref{atmpars}. The black solid line represents the combined spectrum and the white solid lines show the individual exposures.

\section{Radial Velocities}
\label{radvel}

The NEID radial velocities and associated uncertainties listed in Table~\ref{rvtab} were calculated using the cross-correlation function \citep[CCF;][]{baranne1996} method. These values were provided as ``Level 2'' products from the NEID DRP. 
We identified one NEID measurement, taken on 2023 October 26, for which the reported RV was highly discrepant ($\Delta$RV$\sim700$ m~s$^{-1}$) from the remainder of the time series. The reported measurement precision for this single high-velocity point is more than 4 times smaller than that of the next most precise measurement for this star, and 8 times smaller than the measurement precision for observations with comparable S/N. This is unrealistically precise and points to a failure of the RV fitting algorithm as the cause of the discrepant velocity. We re-inspected the data reduction for this night to determine the source of the outlier, and we found that it could be traced to a poor CCF fit for echelle order 127. This was in turn caused by a 100-fold change in flux spanning a single resolution element at 4822\AA, likely due to a cosmic ray. We exclude this order and recompute the RV for this observation. The updated measurement is listed in Table~\ref{rvtab} alongside the pipeline RVs for the rest of the observations and additional information about the data. 

From the literature, we collected 61 radial velocity measurements for \rave, spanning 31 years (from 1984 to 2015). Details on this compilation, including references, can be found in Table~\ref{rvlit}. 
Figure~\ref{rvfig} shows, on the upper panels, the radial velocity measurements for \rave\, taken from the literature (left) and from this work (right). The bottom panel shows in more detail the NEID measurements and their uncertainties. Also shown are the median value and the median absolute deviation (MAD), listed in Table~\ref{starinfo}. 
In comparison with the literature compilation, the NEID values have remarkably small scatter (MAD~=~16~\msec) and individual uncertainties, reaching as low as 15.3~\msec. It must be noted that the work of \citet{hansen2016} reached a 51~\msec\, standard deviation from 18 radial velocity measurements of \rave, using the  FIES spectrograph at the 2.5-meter Nordic Optical Telescope on La Palma, Spain. To date, the NEID radial velocity measurements presented in this work are the most precise ever taken for an EMP star.

\section{Atmospheric Parameters and Chemical Abundances}
\label{atmpars}

\subsection{Atmospheric Parameters}

The stellar atmospheric parameters for \rave\, (\teff, \logg, \metal, and $\xi$) were determined by a combination of photometric and spectroscopic methods, as described below.
The \teff\, was calculated from the color-\teff-\metal\, relations derived by \citet{mucciarelli21}. The same procedure outlined in \citet{roederer2018} was used, drawing 10$^5$ samples for magnitudes, reddening, and metallicity. The $G$, $BP$, and $RP$ magnitudes were gathered from the third data release of the Gaia mission \citep[DR3;][]{gaia23dr3} and the $K$ magnitude from 2MASS \citep{skrutskie2006}. The weighted mean of the median temperatures for each input color ($BP-RP$, $BP-G$, $G-RP$, $BP-K$, $RP-K$, and $G-K$) produces the final value of \teff=$5351\pm51$~K, which is about 80~K cooler than the estimate from \citet{ito2013}, derived from the relations in \citet{casagrande2010} using the $V-K_s$ color.
The \logg\, was calculated from Equation~1 in \citet{roederer2018}, drawing 10$^5$ samples from the input parameters listed in Table~\ref{starinfo}. The final value for \rave\, (\logg=$3.12\pm0.07$) is taken as the median of those calculations with the uncertainty given by their standard deviation. \citet{ito2013} adopted \logg=$3.40\pm0.30$ from the ionization equilibrium of \ion{Fe}{1}/\ion{Fe}{2} and \ion{Ti}{1}/\ion{Ti}{2} abundances.

The metallicity for \rave\, was determined spectroscopically from the equivalent widths (EWs) of 123 \ion{Fe}{1} absorption features in the NEID spectrum, by fixing the \teff\ and \logg\ determined above. Table~\ref{eqw} lists the lines analyzed in this work, their measured EWs, and the derived chemical abundances. The EWs were obtained by fitting Gaussian profiles to the observed features using standard \texttt{NOIRLab IRAF} routines. The \metal\, was then calculated using the 2017 version of the \texttt{MOOG}\footnote{\href{https://github.com/alexji/moog17scat}{https://github.com/alexji/moog17scat}} code \citep{sneden1973}, employing one-dimensional plane-parallel model atmospheres with no overshooting \citep{castelli2004}, assuming local thermodynamic equilibrium (LTE). Finally, the microturbulent velocity ($\xi$) was determined by minimizing the trend between the \ion{Fe}{1} abundances and their reduced equivalent width ($\log(\rm{EW}/\lambda$)). The final atmospheric parameters and uncertainties for \rave\, are listed in Table~\ref{starinfo}, as well as the values from \citet{ito2013} for comparison.

\begin{deluxetable}{lr@{}r@{}r@{}r@{}rr}[!ht] 
\tabletypesize{\tiny}
\tabletypesize{\footnotesize}
\tablewidth{0pc}
\tablecaption{\label{eqw} Atomic Data and Derived Abundances}
\tablehead{
\colhead{Ion}&
\colhead{$\lambda$}&
\colhead{$\chi$} &
\colhead{$\log\,gf$}&
\colhead{$EW$}&
\colhead{$\log\epsilon$\,(X)}&
\colhead{$\Delta$}\\
\colhead{}&
\colhead{({\AA})}&
\colhead{(eV)} &
\colhead{}&
\colhead{(m{\AA})}&
\colhead{}&
\colhead{NLTE}}
\startdata
\ion{Li}{1}  & 6707.80 & 0.00    &    0.17 &     syn &    0.86 &     0.006 \\
CH           & 4313.00 & \nodata & \nodata &     syn &    5.87 &   \nodata \\
\ion{Na}{1}  & 5889.95 & 0.00    &    0.11 &     syn &    2.63 &  $-$0.107 \\
\ion{Na}{1}  & 5895.92 & 0.00    & $-$0.19 &     syn &    2.58 &  $-$0.052 \\
\ion{Mg}{1}  & 3829.35 & 2.71    & $-$0.23 &   91.12 &    4.13 &     0.197 \\
\ion{Mg}{1}  & 3832.30 & 2.71    &    0.25 &  121.40 &    4.17 &     0.148 \\
\ion{Mg}{1}  & 3838.29 & 2.72    &    0.47 &  137.95 &    4.17 &     0.126 \\
\ion{Mg}{1}  & 4702.99 & 4.33    & $-$0.44 &   12.49 &    4.37 &     0.150 \\
\ion{Mg}{1}  & 5172.68 & 2.71    & $-$0.36 &   97.37 &    4.17 &     0.170 \\
\ion{Mg}{1}  & 5183.60 & 2.72    & $-$0.17 &  109.89 &    4.18 &     0.154 \\
\nodata      & \nodata & \nodata & \nodata & \nodata & \nodata &   \nodata \\ 
\nodata      & \nodata & \nodata & \nodata & \nodata & \nodata &   \nodata \\ 
\enddata
    \tablecomments{The complete list of absorption features and literature references are given in Table~\ref{eqwl}.}
\end{deluxetable}

\begin{deluxetable}{lcrrrcrc}[!ht] 
\tabletypesize{\small}
\tabletypesize{\footnotesize}
\tabletypesize{\scriptsize}
\tablewidth{0pc}
\tablecaption{LTE Abundances for Individual Species \label{abund}}
\tablehead{
\colhead{Species}                     & 
\colhead{$\log\epsilon_{\odot}$\,(X)} & 
\colhead{$\log\epsilon$\,(X)}         & 
\colhead{$\mbox{[X/H]}$}              & 
\colhead{$\mbox{[X/Fe]}$}             & 
\colhead{$\sigma$}                    & 
\colhead{$N$}                         &
\colhead{Ref.}}
\startdata 
 \ion{Li}{1}       &  1.05 &     0.86 &  $-$0.19 &     3.77 &    0.10 &   1 & 1 \\ 
 \ion{C}{0}\vv     &  8.43 &     5.87 &  $-$2.56 &     1.40 &    0.10 &   1 & 1 \\ 
 \ion{N}{0}        &  7.83 &     3.90 &  $-$3.93 &     0.03 &    0.20 &   1 & 2 \\ 
 \ion{O}{0}        &  8.69 &     6.25 &  $-$2.44 &     1.52 &    0.20 &   2 & 2/3 \\ 
 \ion{Na}{1}       &  6.24 &     2.61 &  $-$3.63 &     0.33 &    0.03 &   2 & 1 \\ 
 \ion{Mg}{1}       &  7.60 &     4.25 &  $-$3.35 &     0.61 &    0.12 &   8 & 1 \\ 
 \ion{Al}{1}       &  6.45 &     1.94 &  $-$4.51 &  $-$0.55 &    0.10 &   1 & 1 \\ 
 \ion{Si}{1}       &  7.51 &     3.90 &  $-$3.61 &     0.35 &    0.10 &   1 & 1 \\ 
 \ion{P}{1}        &  5.41 &     1.04 &  $-$4.37 &  $-$0.41 &    0.21 &   3 & 4 \\ 
 \ion{S}{1}        &  7.12 &     3.37 &  $-$3.75 &     0.21 &    0.41 &   3 & 4 \\ 
 \ion{Ca}{1}       &  6.34 &     2.73 &  $-$3.61 &     0.35 &    0.11 &   7 & 1 \\ 
 \ion{Sc}{2}       &  3.15 &  $-$0.61 &  $-$3.76 &     0.20 &    0.10 &   1 & 1 \\ 
 \ion{Ti}{1}       &  4.95 &     1.39 &  $-$3.56 &     0.40 &    0.07 &   7 & 1 \\ 
 \ion{Ti}{2}       &  4.95 &     1.19 &  $-$3.76 &     0.20 &    0.04 &  10 & 1 \\ 
 \ion{V}{2}        &  3.93 &  $-$0.15 &  $-$4.08 &  $-$0.12 &    0.15 &   2 & 2 \\ 
 \ion{Cr}{1}       &  5.64 &     1.28 &  $-$4.36 &  $-$0.40 &    0.03 &   2 & 1 \\ 
 \ion{Cr}{2}       &  5.64 &     2.06 &  $-$3.58 &     0.38 &    0.21 &   2 & 4 \\ 
 \ion{Mn}{1}       &  5.43 &     0.40 &  $-$5.03 &  $-$1.07 &    0.03 &   2 & 1 \\ 
 \ion{Mn}{2}       &  5.43 &     0.75 &  $-$4.68 &  $-$0.72 &    0.02 &   3 & 2 \\ 
 \ion{Fe}{1}       &  7.50 &     3.54 &  $-$3.96 &     0.00 &    0.09 & 123 & 1 \\ 
 \ion{Fe}{2}       &  7.50 &     3.57 &  $-$3.93 &     0.03 &    0.09 &   6 & 1 \\ 
 \ion{Co}{1}       &  4.99 &     1.47 &  $-$3.52 &     0.44 &    0.07 &   4 & 1 \\ 
 \ion{Co}{2}       &  4.99 &     1.55 &  $-$3.44 &     0.52 &    0.23 &   4 & 4 \\ 
 \ion{Ni}{1}       &  6.22 &     2.34 &  $-$3.88 &     0.08 &    0.05 &   2 & 1 \\ 
 \ion{Ni}{2}       &  6.22 &     2.09 &  $-$4.13 &  $-$0.17 &    0.21 &   3 & 4 \\ 
 \ion{Cu}{1}       &  4.19 &  $-$1.04 &  $-$5.23 &  $-$1.27 &    0.15 &   1 & 2 \\ 
 \ion{Zn}{2}       &  4.56 &     0.33 &  $-$4.23 &  $-$0.27 &    0.24 &   1 & 4 \\ 
 \ion{Sr}{2}       &  2.87 &  $-$1.41 &  $-$4.28 &  $-$0.32 &    0.10 &   2 & 1 \\ 
 \ion{Ba}{2}       &  2.18 &  $-$2.53 &  $-$4.71 &  $-$0.75 &    0.10 &   2 & 1 \\ 
 \ion{Eu}{2}       &  0.52 & $<-$2.62 & $<-$3.14 & $<$~0.82 & \nodata &   1 & 1 \\ 
\enddata
\tablerefs{(1) WIYN/NEID (This work); (2) Subaru/HDS \citep{ito2013}; (3) HST/STIS \citep{placco2014b}; (4) HST/COS \citep{roederer2016}.}
\tablenotetext{a}{No evolutionary corrections from \citet{placco2014c}.}
\end{deluxetable}

\begin{figure*}
 \includegraphics[width=1\linewidth]{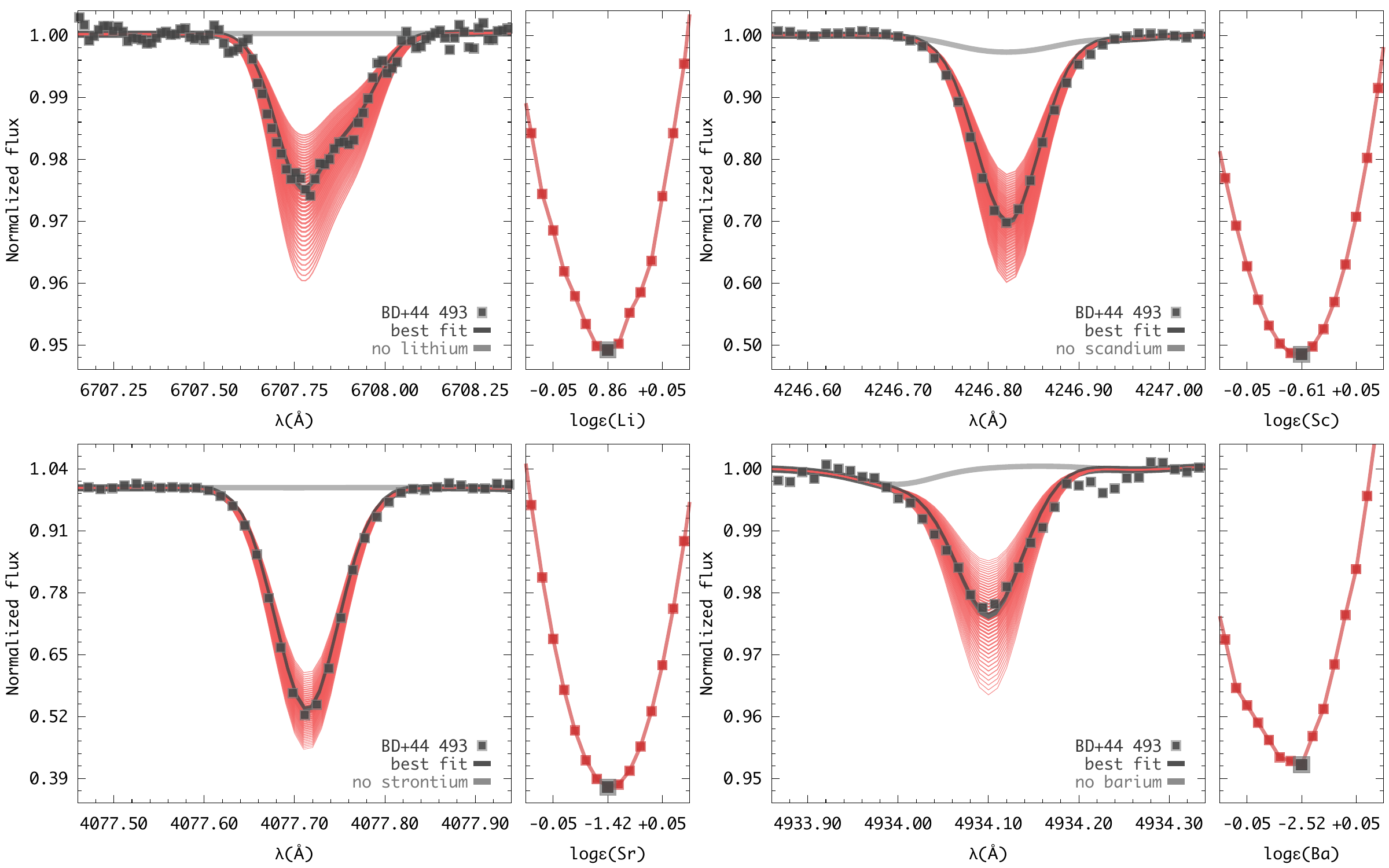}
     \caption{Chemical abundance determination for lithium (upper left), scandium (upper right), strontium (lower left), and barium (lower right). The points represent the NEID combined spectrum, the black lines are the best fits, and the red lines are synthetic spectra with varying abundances for each element. Also shown are synthetic spectra without the presence of each element (gray line). The right panels show the $\chi^2$ minimization routine for each synthetic spectra shown in the left panels. The x-axis is centered on the best-fit value (represented by the gray square) and also shown are $\pm0.05$~dex marks for reference.}
 \label{moog}
\end{figure*}

\subsection{Chemical Abundances}

We measured 182 absorption features for 16 elements in the NEID combined spectra, from 3743\AA\, (\ion{Fe}{1}) to 8806\AA\, (\ion{Mg}{1}). Abundances were calculated from equivalent-width analysis and spectral synthesis\footnote{Equivalent-width analysis is used for mostly isolated spectral features, while spectral synthesis is important for blended features, molecules, and when line broadening by isotopic shifts and hyperfine splitting structure has to be taken into account.}, also using the 2017 version of \texttt{MOOG}. The atomic data, EW values where applicable, and abundances for these features are listed in Table~\ref{eqw}. Abundances determined from spectral synthesis have a {\emph{syn}} label on the EW column. The linelists for the spectral synthesis were generated using the \texttt{linemake} code\footnote{\href{https://github.com/vmplacco/linemake}{https://github.com/vmplacco/linemake}} \citep{placco2021}. Logarithmic number abundances ($\log\epsilon$(X)) and abundance ratios (\abund{X}{H} and \xfe{X}), were calculated adopting the solar photospheric abundances ($\log\epsilon_{\odot}$\,(X)) from \citet{asplund2009}. Average abundances and the number of lines measured ($N$) for each element are given in Table~\ref{abund}. The $\sigma$ values represent the standard error of the mean. For the elements where $N=1$, the uncertainties were estimated by minimizing the residuals between the NEID data and a set of synthetic spectra, as described in detail below.

\begin{figure*}
 \includegraphics[width=1\linewidth]{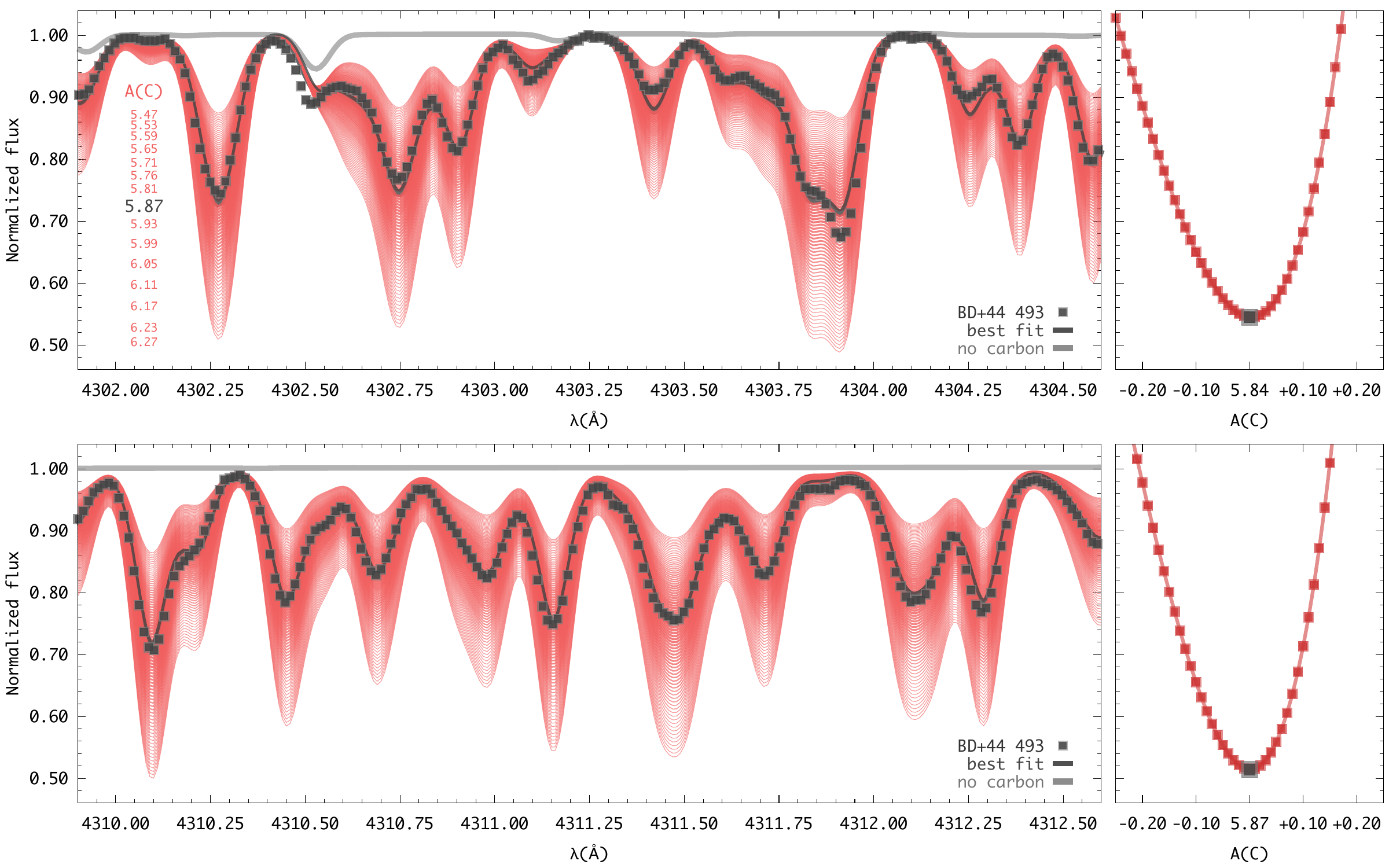}
     \caption{Left panels: Carbon abundance determination via spectral synthesis for two 2.5\AA\, wide sections of the CH G-band. The points represent the NEID combined spectrum, the black line is the best fit, and the red lines are synthetic spectra with varying carbon abundances. Also shown is a synthetic spectrum without carbon (gray line). Right panels: $\chi^2$ minimization routine for each synthetic spectra shown in the left panels. The x-axis is centered on the best-fit value (represented by the gray square) and also shown are $\pm0.1$ and $\pm0.2$~dex marks for reference.}
 \label{carbon}
\end{figure*}

The upper left panel of Figure~\ref{moog} shows the abundance determination for the \ion{Li}{1} doublet at 6707\AA\, absorption feature found in the atmosphere of \rave. The black squares represent the NEID combined spectrum, the black line is the best fit, and the red lines are synthetic spectra with varying Li abundances (in steps of 0.01~dex). Also shown is a synthetic spectrum without Li (gray line). The right panel illustrates the search for the best-fit abundance value and the uncertainty estimation. We calculate the $\chi^2$ for each synthetic spectra shown in the left panels and find the global minimum as the most probable value. The x-axis on the right panel is centered on the best-fit value (represented by the gray square) and also shown are $\pm0.05$~dex marks for reference. For the synthesis, we assume that only the $^7$Li isotope contributes to the line strength, without any contribution from $^6$Li. The best-fit abundance is \eps{Li}=0.86, which is 0.14~dex lower than the value from \citet{ito2013}. This offset is driven by the differences in stellar parameters, particularly \teff. We generated synthetic spectra using the NEID data and a model atmosphere with the parameters from \citet{ito2013} and were able to reproduce their \eps{Li} value.

For the carbon abundance determination, we synthesized the CH G-band in six 2.5\AA-wide regions in the 4290-4330\AA\, range and determined the best-fit value for each region using the procedure outlined above. Two examples are shown in Figure~\ref{carbon}. It is possible to see a remarkably good agreement between the high S/N NEID data ($\sim400$ per pixel at 4300\AA) and the synthetic spectra. 
We attempted to determine the $^{12}$C/$^{13}$C carbon isotopic ratio from the 4217-4237\AA\, region. However, the $^{13}$CH features were too weak for meaningful detection. Based on the synthetic data for a few different isotopic ratios, we set $^{12}$C/$^{13}$C=49 for all the syntheses performed in this work, which is in agreement with the $^{12}$C/$^{13}$C$>30$ value from \citet{ito2013}.
Finally, due to its position at the base of the giant branch with \logg=3.12, \rave\, has not experienced significant carbon depletion in its atmosphere, so there is no evolutionary correction to the carbon abundance according to \citet{placco2014c}.
The adopted carbon abundance, \eps{C}=5.87, is the mean best-fit value for the six regions. We conservatively set the uncertainty as 0.10~dex.

For the light elements ($Z<30$), we measured abundances from spectral synthesis for the \ion{Na}{1} doublet (5889 and 5895\AA), \ion{Al}{1} (3961\AA), \ion{Si}{1} (3905\AA), and \ion{Sc}{2} (4246\AA\, - see the upper right panel on Figure~\ref{moog}). The remaining elements listed in Table~\ref{abund} had their abundances measured through EW analysis. A comparison between the EW measurements from this work and \citet{ito2013} is provided in Section~\ref{itocomp}.

\begin{figure}
 \includegraphics[width=1.0\linewidth]{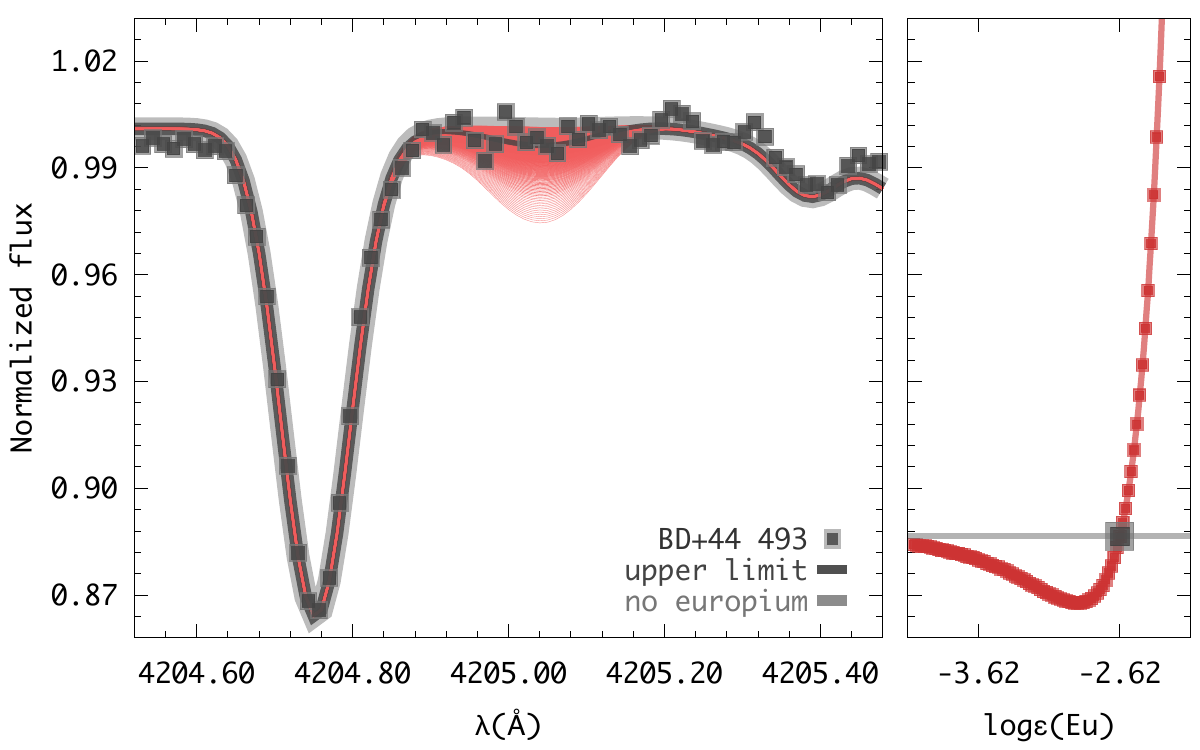}
     \caption{Upper limit estimation for the europium abundance in \protect\rave, using the absorption feature at 4205\AA. Points and lines are similar to the ones in Figures~\ref{moog} and \ref{carbon}. See text for further details.}
 \label{upperlim}
\end{figure}

We measured the chemical abundances of two heavy elements in the NEID data: strontium (4077\AA\, and 4215\AA) and barium (4554\AA\, and 4934\AA). The bottom panels of Figure~\ref{moog} show the syntheses for two of those absorption features. For both Sr and Ba, the abundances for each line agree within 0.02~dex. A similar $\chi^2$ minimization approach was used to estimate an upper limit for the europium abundance, using the 4205\AA\, feature. The results are shown in Figure~\ref{upperlim}. We chose to set a conservative upper limit as the \eps{Eu} for which the $\chi^2$ matches the value for the spectra without Eu on the opposite side of the global minimum, as shown in the right panel of Figure~\ref{upperlim}. Simply adopting the minimum $\chi^2$ could lead to an underestimation of the upper limit, as the $\chi^2$ becomes less sensitive to changes in the abundance for very low \eps{Eu}. The final estimate, \eps{Eu}$<-2.62$, is about 16 times lower than the upper limit set by \citet{placco2014b} and a factor of two higher than the one from \citet{ito2013}.

\subsection{Complementary chemical abundances from the Subaru and HST spectra}

In addition to the abundances measured from the NEID spectra, we have re-determined abundances for selected elements using the updated stellar parameters from this work and data from \citet{ito2013,aoki2015} (Subaru/HDS), \citet{placco2014b} (HST/STIS), and \citet{roederer2016} (HST/COS). The final values are listed in Table~\ref{abund} with notes that indicate their provenance.

The nitrogen abundance for \rave\, was determined by synthesizing the NH molecular feature around 3360\AA\, present in the Subaru/HDS spectrum. We were able to reproduce the abundance from \citet{ito2013} when using their stellar parameters, and the final value, \eps{N}=3.90, was determined using the stellar parameters calculated from the NEID data.
From the HST/STIS data, we measured an abundance of oxygen (\eps{O}=6.21) from several OH molecular features in the 2965-2972\AA\, range. This value is 0.14~dex smaller than the one reported in \citet{placco2014b}. From the Subaru/HDS data, we synthesized other OH features around 3140\AA\, with the best fit at \eps{O}=6.29. The final O abundance reported in Table~\ref{abund} is the mean of the two values mentioned above, \eps{O}=6.25.
Also from the Subaru/HDS spectra, we obtained abundances for vanadium (3545\AA\, and 3592\AA; \eps{V}=$-$0.15), manganese (3441\AA, 3460\AA, and 3488\AA; \eps{Mn}=0.75), and copper  (3274\AA; \eps{Cu}=$-$1.04).

Following the work of \citet{roederer2016}, we used their EW values measured from the HST/COS data\footnote{\citeauthor{roederer2016} measured the EW values by fitting the lines with a convolution of a Gaussian and the COS line-spread function.} to re-determine several abundances for \rave\, with the updated stellar parameters: phosphorus (three lines; \eps{P}=1.04), sulphur (three lines; \eps{S}=3.37), \ion{Cr}{2} (two lines; \eps{Cr}=2.06), \ion{Co}{2} (four lines; \eps{Co}=1.55), \ion{Ni}{2} (three lines; \eps{Ni}=2.09), and zinc (one line; \eps{Zn}=0.33) in the 1800-2150\AA\, wavelength range. The $\sigma$ values for these abundances were taken from Table~2 in \citet{roederer2016}.

\subsection{Non-LTE Corrections}

Non-LTE (NLTE) corrections were obtained for 159 absorption features in the spectrum of \rave, using the following databases: INSPECT\footnote{\href{http://www.inspect-stars.com/}{http://www.inspect-stars.com/}} (\ion{Li}{1} and \ion{Na}{1}), \citet{nordlander2017b} (\ion{Al}{1}), and MPIA NLTE\footnote{\href{https://nlte.mpia.de/}{https://nlte.mpia.de/}} (\ion{Mg}{1}, \ion{Si}{1}, \ion{Ca}{1}, \ion{Ti}{1}, \ion{Ti}{2}, \ion{Cr}{1}, \ion{Mn}{1}, \ion{Fe}{1}, \ion{Fe}{2}, and \ion{Co}{1}) -- the NLTE corrections for individual lines and literature references are given in Table~\ref{eqwl}. Average NLTE abundances, abundance ratios, and $\sigma$ values are shown in Table~\ref{abundn}. The average NLTE corrections range from $-0.08$ for \ion{Na}{1} to $+0.95$ for \ion{Mn}{1}, with notably high corrections also for \ion{Cr}{1} and \ion{Al}{1} ($+0.81$ and $+0.60$, respectively).
%
%

\begin{deluxetable}{lcrrrcr}[!ht] 
\tabletypesize{\small}
\tabletypesize{\footnotesize}
\tablewidth{0pc}
\tablecaption{NLTE Abundances for Individual Species \label{abundn}}
\tablehead{
\colhead{Species}                     & 
\colhead{$\log\epsilon_{\odot}$\,(X)} & 
\colhead{$\log\epsilon$\,(X)}         & 
\colhead{$\mbox{[X/H]}$}              & 
\colhead{$\mbox{[X/Fe]}$}             & 
\colhead{$\sigma$}                    & 
\colhead{$N$}}
\startdata %
 \ion{Li}{1}    &  1.05 &    0.87 & $-$0.18 &    3.59 &    0.10 &   1 \\ 
 \ion{Na}{1}    &  6.24 &    2.60 & $-$3.64 &    0.13 &    0.04 &   2 \\ 
 \ion{Mg}{1}    &  7.60 &    4.41 & $-$3.19 &    0.58 &    0.12 &   8 \\ 
 \ion{Al}{1}    &  6.45 &    2.54 & $-$3.91 & $-$0.14 &    0.10 &   1 \\ 
 \ion{Si}{1}    &  7.51 &    4.01 & $-$3.50 &    0.27 &    0.10 &   1 \\ 
 \ion{Ca}{1}    &  6.34 &    2.88 & $-$3.46 &    0.31 &    0.14 &   6 \\ 
 \ion{Ti}{2}    &  4.95 &    1.27 & $-$3.68 &    0.09 &    0.04 &  10 \\ 
 \ion{Cr}{1}    &  5.64 &    2.09 & $-$3.55 &    0.22 &    0.05 &   2 \\ 
 \ion{Mn}{1}    &  5.43 &    1.35 & $-$4.08 & $-$0.31 &    0.05 &   2 \\ 
 \ion{Fe}{1}    &  7.50 &    3.73 & $-$3.77 &    0.00 &    0.10 & 120 \\ 
 \ion{Fe}{2}    &  7.50 &    3.57 & $-$3.93 & $-$0.16 &    0.10 &   6 \\ 
\enddata
\end{deluxetable}

\vspace{-1cm}

\subsection{Comparison with \citet{ito2013}} 
\label{itocomp}

\begin{figure}[!ht]
 \includegraphics[width=1\linewidth]{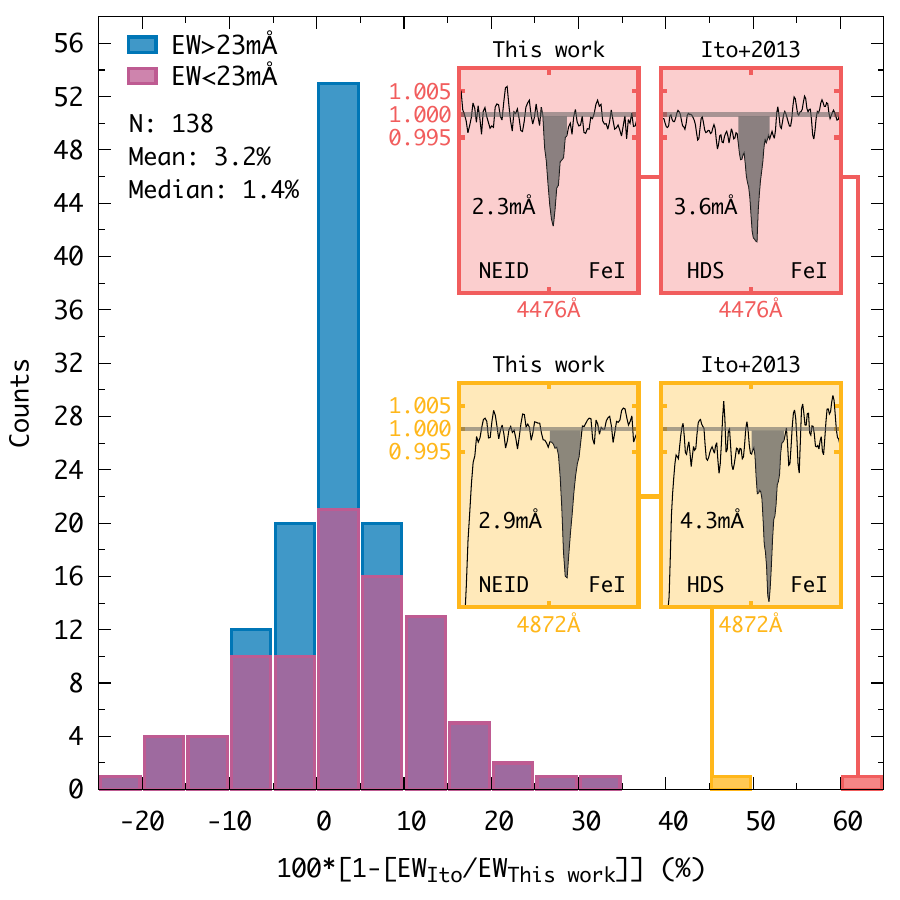}
     \caption{Distribution of percent differences in equivalent width (in m\AA) for the 138 absorption features in common with the \citet{ito2013} analysis. The mean and median differences are shown in the upper left. The red and yellow insets show the lines with the largest percent differences, highlighting the continuum placement in each case (NEID on the left, HDS on the right). See text for further details.}
 \label{itoHist}
\end{figure}

There are 138 absorption features that were measured by both \citet{ito2013} and this work: \ion{Fe}{1} (106 lines), \ion{Mg}{1} and \ion{Ti}{2} (7 lines each), \ion{Ti}{1} and \ion{Fe}{2} (5 lines each), \ion{Co}{1} (4 lines), and 2 lines each for \ion{Cr}{1} and \ion{Mn}{1}. Figure~\ref{itoHist} shows the distribution of percent differences in the measured equivalent widths. There is an overall good agreement (3.2\% mean and 1.4\% median for the 138 lines) between the measurements, especially for EW$>$23m\AA\footnote{The 23m\AA\, value was chosen based on the EW distribution for illustration purposes only.} (blue bars on the stacked histogram). This is expected since both Subaru/HDS and WYIN/NEID data for \rave\, are of high S/N values. From the 138 common features, 86 are at $\lambda\leq4500$\AA, 106 are at $\lambda\leq5000$\AA\, and only 6 with $\lambda\geq5500$\AA.

For weaker absorption features, continuum placement has a larger effect on the EW measurements. Two examples (\ion{Fe}{1} lines at 4476\AA\, and 4872\AA\, with S/N per pixel in the NEID data of 396 and 538, respectively) that show the largest differences are highlighted in the colored insets of Figure~\ref{itoHist}. The left panels show the continuum placement and EW values calculated using the NEID data, and the right panels have the continuum level and EW values measured by \citet{ito2013}, also using the Subaru/HDS data. It is possible to see that a difference of about 0.002 in the overall continuum placement can drastically change the abundance of weak absorption lines, which highlights the importance of having high S/N data.

\subsection{Systematic Uncertainties}

Systematic uncertainties due to changes in the atmospheric parameters were quantified for the elements with $Z\leq28$ and abundances determined by equivalent analysis\footnote{We calculated the EW values for the \ion{Li}{1}, \ion{Na}{1}, \ion{Al}{1}, \ion{Si}{1}, \ion{Sc}{2}, and \ion{Ni}{1} lines for this exercise. The abundances reported in Table~\ref{abund} were determined from spectral synthesis.}, following \citet{placco2013,placco2023b}.  Table~\ref{sys} shows the abundance variations when the atmospheric parameters are changed within the quoted values ($+$150~K for \teff, $+$0.25~dex for \logg, and $+$0.25~km\,s$^{-1}$ for $\xi$). The $\sigma$ values are taken directly from Table~\ref{abund}. The total uncertainty ($\sigma_{\rm tot}$) for each element is calculated from the quadratic sum of the individual error estimates. For the elements not listed in Table~\ref{sys}, the $\sigma$ values from Table~\ref{abund} should be used.

\begin{deluxetable}{@{}lrrrrr@{}}[!ht]
\tabletypesize{\small}
\tabletypesize{\footnotesize}
\tablewidth{0pc}
\tablecaption{Systematic Abundance Uncertainties for \protect\rave \label{sys}}
\tablehead{
\colhead{Elem}&
\colhead{$\Delta$\teff}&
\colhead{$\Delta$\logg}&
\colhead{$\Delta\xi$}&
\colhead{$\sigma$}&
\colhead{$\sigma_{\rm tot}$\vv}\\
\colhead{}&
\colhead{$+$150\,K}&
\colhead{$+$0.25 dex}&
\colhead{$+$0.25 km/s}&
\colhead{}&
\colhead{}}
\startdata
\ion{Li}{1} &    0.14 &    0.01 &    0.00 &    0.10 &    0.17 \\
\ion{Na}{1} &    0.14 & $-$0.01 & $-$0.03 &    0.03 &    0.15 \\
\ion{Mg}{1} &    0.13 & $-$0.04 & $-$0.04 &    0.12 &    0.19 \\
\ion{Al}{1} &    0.15 &    0.01 & $-$0.02 &    0.10 &    0.18 \\
\ion{Si}{1} &    0.15 & $-$0.01 & $-$0.07 &    0.10 &    0.19 \\
\ion{Ca}{1} &    0.11 &    0.00 & $-$0.02 &    0.11 &    0.16 \\
\ion{Sc}{2} &    0.11 &    0.09 & $-$0.02 &    0.10 &    0.17 \\
\ion{Ti}{1} &    0.16 &    0.01 & $-$0.01 &    0.07 &    0.18 \\
\ion{Ti}{2} &    0.09 &    0.10 & $-$0.03 &    0.04 &    0.14 \\
\ion{Cr}{1} &    0.18 &    0.01 & $-$0.01 &    0.03 &    0.18 \\
\ion{Mn}{1} &    0.20 &    0.01 &    0.00 &    0.03 &    0.20 \\
\ion{Fe}{1} &    0.16 &    0.00 & $-$0.04 &    0.09 &    0.19 \\
\ion{Fe}{2} &    0.03 &    0.09 & $-$0.01 &    0.09 &    0.13 \\
\ion{Co}{1} &    0.18 &    0.01 & $-$0.01 &    0.07 &    0.19 \\
\ion{Ni}{1} &    0.19 &    0.01 & $-$0.02 &    0.05 &    0.20 \\
\enddata
\tablenotetext{a}{Calculated from the quadratic sum of the individual error estimates.}
\end{deluxetable}

\section{The Chemo-dynamical Nature of \protect\rave}
\label{chemod}

\subsection{The Light-element Abundance Pattern}

\rave\, is believed to be a bonafide second-generation star, formed from a gas cloud polluted by a single Pop. III supernova explosion. The light-element ($Z\leq30$) abundance pattern of \rave\, has been compared with a variety of different stellar progenitors, with the best agreement coming from the so-called ``faint supernovae'' associated with the evolution of massive metal-free stars in the early universe. Further constraints on the nature of \rave\, also come from its low magnesium-to-carbon ratio, \abund{Mg}{C}=$-$0.79, which places it in the ``mono-enriched'' regime proposed by \citet{hartwig2018}.

\begin{figure*}[!ht]
 \includegraphics[width=0.5\linewidth]{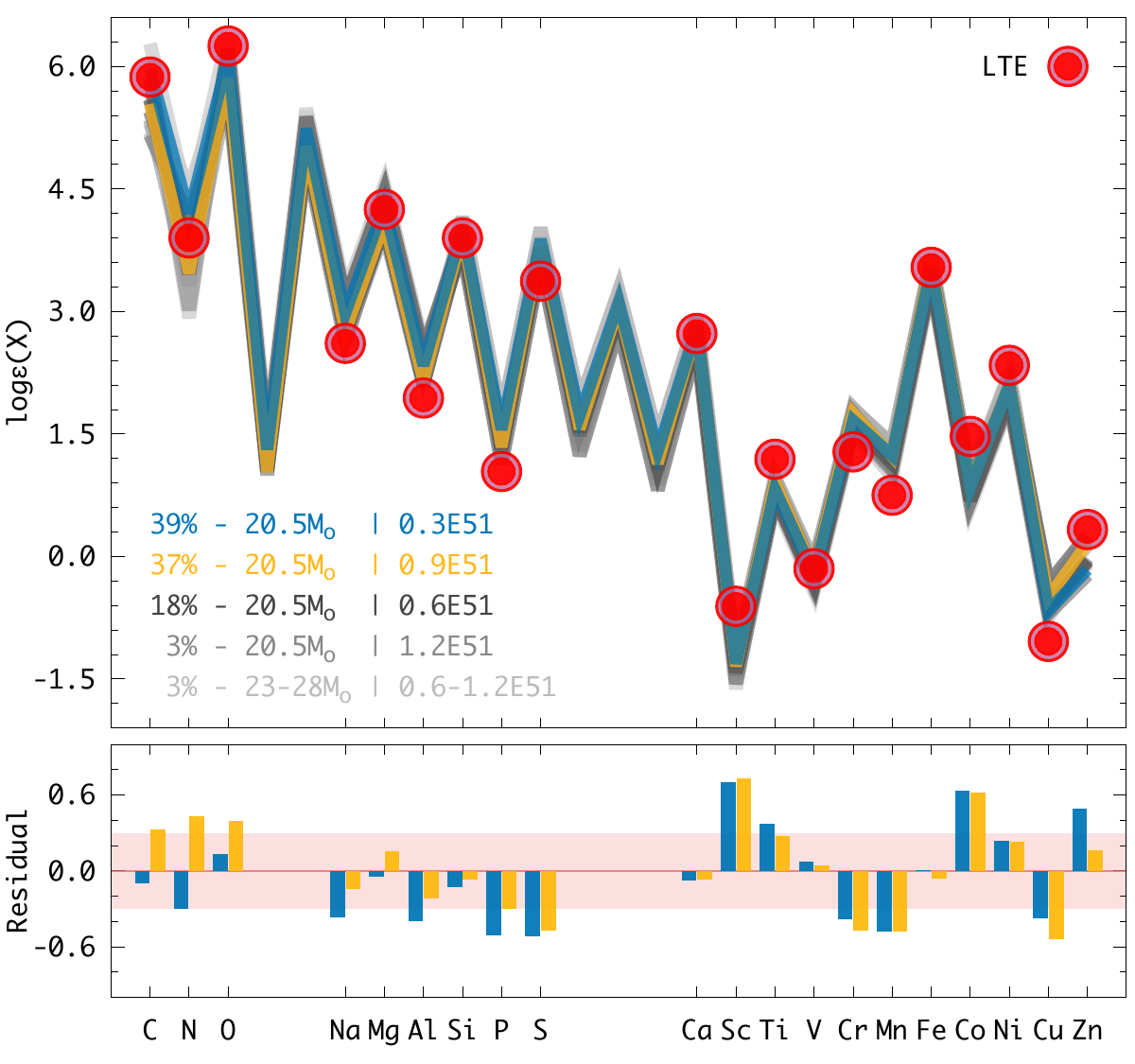}
 \includegraphics[width=0.5\linewidth]{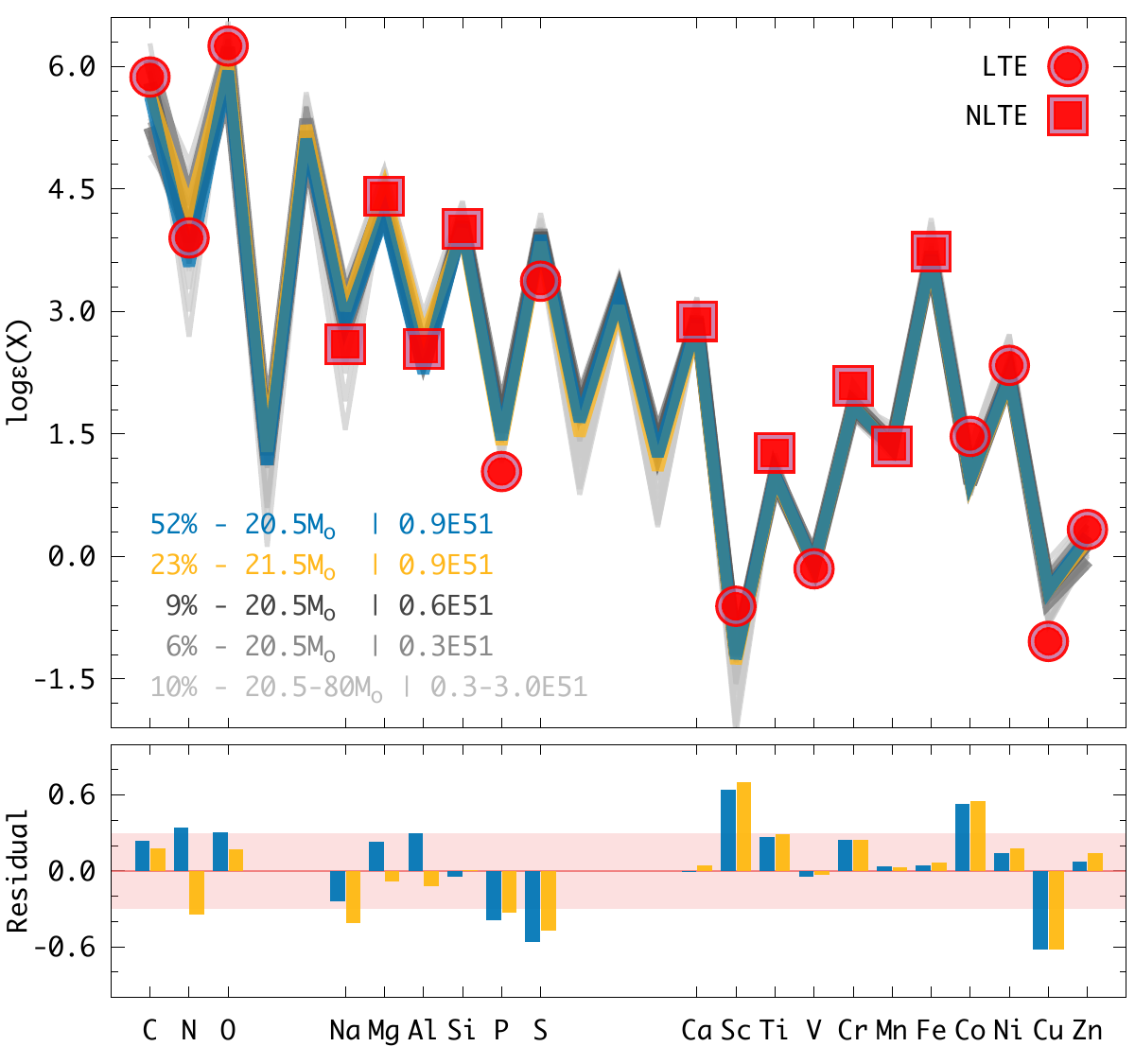}
     \caption{Upper panels: \rave\, light-element abundances (filled circles represent LTE values - left panels - and filled squares represent NLTE values - right panels) compared with yields from metal-free supernova models (solid lines). The labels show each model's progenitor masses, explosion energies, and their percentage occurrence among the 10,000 resamples of the \rave\, abundances. Lower panel: residuals between the observations and the two best-fit models. A $\pm0.3$ dex shaded area is shown for reference.}
 \label{starfit}
\end{figure*}

Using the models from \citet{umeda2005} and \citet{tominaga2007}, \citet{ito2013} found a good agreement when comparing the abundances of \rave\, (16 elements with $6\leq Z \leq 30$) with the yields from a 25\,M$_\odot$, $5.0 \times 10^{51}$\,erg explosion energy mixing-and-fallback supernova. \citet{roederer2016}, comparing abundances for 20 elements (including P, S, and Zn) with the yields from \citet{heger2010}, found a best-fit faint supernova model with 20.5\,M$_\odot$ and $0.6 \times 10^{51}$\,erg.  \citet{placco2016b}, with a subset\footnote{\citeauthor{placco2016b} used the highest number of elements in common that were available for a sample of 12 stars with \metal$\lesssim-4$ in the literature.} of the abundances from \citeauthor{roederer2016}, found their best-fit model to be 21.5\,M$_\odot$ and $0.3 \times 10^{51}$\,erg.

With the abundances determined in this work using the updated stellar parameters for \rave, we were able to repeat the exercise outlined in \citet{placco2020,placco2021b} and \citet{mardini2022} by using the \texttt{starfit}\footnote{\href{https://starfit.org/}{https://starfit.org/}} tool, which contains Pop. III supernova nucleosynthesis yields from \citet{heger2010}. We re-sampled the chemical abundance pattern of \rave\, 10,000 times, assuming Gaussian distributions for each element in Table~\ref{abund}, centered in the \eps{X}\footnote{For elements with abundances for neutral and ionized species we used the one with the smallest uncertainty (e.g. \ion{Ti}{2} and \ion{Mn}{2}).} values and with dispersion $\sigma$. For the elements with $\sigma<0.1$ in Table~\ref{abund}, we adopted a standard $\sigma=0.2$ to conservatively account for the systematic uncertainties shown in Table~\ref{sys}. 

The results are shown in the left panels of Figure~\ref{starfit}, where the red circles are the measured abundances for \rave\, and the lines represent the various best-fit models, colored by their frequency, mass, and explosion energy labeled in the top panel. A total of 13 different models were found to be the best fit for the 10,000 re-sampled abundance patterns. The bottom panel shows the residuals between the abundances and the two most frequent models. A $\pm0.3$ dex shaded area (representing roughly $2\sigma$ uncertainties for most light elements) is shown for reference. In summary, the best-fit progenitor for 97\% of the 10,000 re-samples has 20.5\,M$_\odot$, with explosion energies within $0.3-1.2 \times 10^{51}$\,erg. Even though the model abundance patterns shown in the top panel appear qualitatively similar, it is important to notice the difference in the residuals, in particular for C, N, and O. These reflect the variation in the explosion energy from the models. The results agree with the best-fit mass found by \citet{roederer2016} but with a lower explosion energy. However, the results with the updated chemical abundances provide a stronger constraint on the progenitor mass.

There are still a few large discrepancies between the best-fit models and the observations. It has been shown that the elements that most constrain the Pop. III progenitor masses for the faint supernova models are carbon and nitrogen \citep{frebel2015b,placco2015,placco2016}. For a few other elements (e.g., Al, Cr, and Mn), the differences can be attributed to NLTE effects. 
Due to the small number of elements with NLTE abundances available for \rave\, (nine elements from Table~\ref{abundn} - excluding \ion{Li}{1} and \ion{Fe}{2}), the \texttt{starfit} matching algorithm is mostly unconstrained, with 1,600 different models found as the best fit for the 10,000 re-samples, as opposed to only 13 for the LTE abundances. 

We combined the LTE and NLTE abundances and repeated the matching exercise, with results shown in the right panels of Figure~\ref{starfit}. The nine NLTE abundances are shown as red squares. The number of unique models found as the best fit for the 10,000 re-samples is 37, about three times higher than the LTE-only case. The most frequent best-fit model is still 20.5\,M$_\odot$ and about 67\% of the re-samples share that mass. The second most frequent model (23\%) has 21.5\,M$_\odot$. In summary, both LTE and LTE$+$NLTE cases confirm the previous findings for the potential stellar progenitor for \rave, supporting the hypothesis that it is a ``mono-enriched'' star. If this is the case, the presence of Sr and Ba in the atmosphere of \rave\, would imply that the progenitor supernova would have been capable of producing such heavy elements. The next step for such analysis is to account for additional NLTE abundances when available and also compare the observed abundances with Pop. III yield prediction from different models.


\subsection{Galactic Orbit and Substructure Membership}
\label{orbits}

\begin{figure}[!ht]
 \includegraphics[width=.93\linewidth]{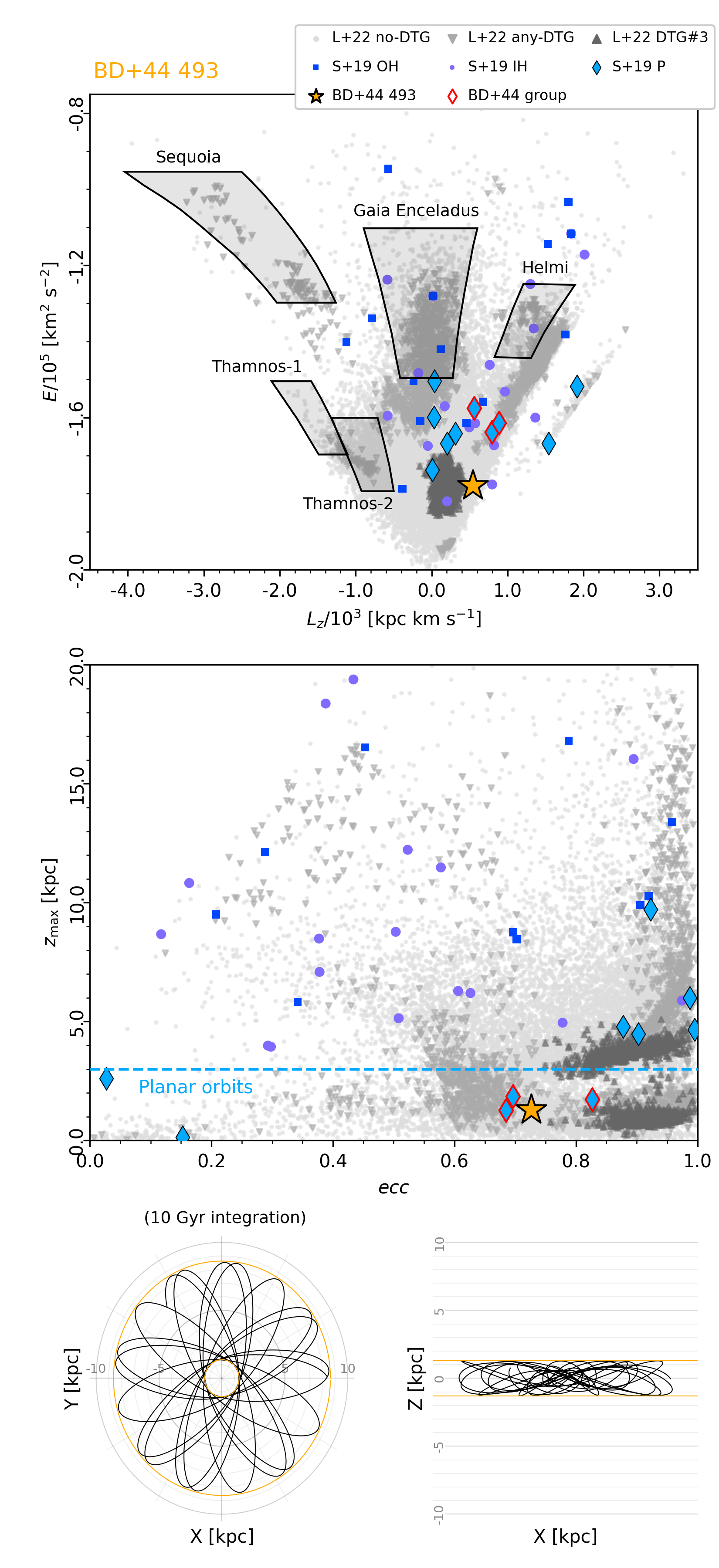}
     \caption{Top: vertical component of the angular momentum and orbital energy of \rave\, (yellow star-shaped symbol), as well as halo substructures defined by \citet[][shaded regions]{Koppelman+2019}, DTGs identified by \citet[][L+22, grey triangles]{Lovdal+2022}, and UMP stars classified by \citet[][S+19]{sestito2019} as outer halo (OH, blue square), inner halo (IH, violet circle) and planar orbits (cyan diamonds). Stars classified in this work as belonging to the same group as \rave\, are highlighted with red-contoured diamonds. Middle: same as above, in the plane of eccentricity ($ecc$) versus maximum distance from Galactic plane ($z_\mathrm{max}$). Bottom: orbital trajectory of \rave\, in the XY-plane (left) and XZ-plane (right).
}
 \label{kin_data}
\end{figure}

We integrated the Galactic orbit of \rave\, using its available 6D kinematic information (position and proper motion from Gaia DR3 -- \citealt{gaia23dr3}; photogeometric distance from \citealt{Bailer-Jones+2021}; and radial velocity from this study). The orbit was integrated for 10 Gyr using the Milky Way Galactic potential from \citet{mcmillan2017}. We adopted $R_0 = 8.21$ kpc and $V_0 = 233.1$ \kmsec. The resulting trajectory is shown in the bottom panels of Figure~\ref{kin_data}. The computed orbital parameters are shown in Table~\ref{starinfo}.
The very small reported uncertainties ($<1\%$) for the kinematical and dynamical parameters are a direct result of the high precision in the astrometric parameters and radial velocity measurements for this bright, nearby star. These uncertainties, however, do not account for systematic errors related to the choice of the Galactic potential used in the orbital integration. Nevertheless, all stars considered in our analysis were integrated using the same Galactic potential, so we do not expect these systematic errors to significantly impact our conclusions.

Despite being an old and extremely metal-poor star, \rave\, displays a distinct planar orbit with high eccentricity. The maximum distance from the Galactic plane ($z_\mathrm{max}$) of just 1.286 kpc is comparable to the scale height of the Milky Way’s thick disk. The star is currently near its apogalactic radius ($R_\mathrm{apo} = 8.623$ kpc), and the estimated perigalactic radius ($R_\mathrm{peri}$) of 1.366 kpc results in an eccentricity ($ecc$) of 0.726. This orbital characteristic was already noted in the literature by \citet[][hereafter \citetalias{sestito2019}]{sestito2019}, who relied on less accurate photometric parallaxes for distance estimation and adopted the \textit{MilkyWay14} potential from \citet{bovy2015} for the Milky Way. In their sample of 42 UMP stars, \citetalias{sestito2019} classified 11 stars as having planar orbits (defined as prograde orbits confined within 3 kpc of the Milky Way plane), showing this characteristic is not particular to \rave.

In the upper panel of Figure~\ref{kin_data} we show the orbital energy and vertical angular momentum of \rave\, (yellow star). The locus associated with members of different halo substructures: Sequoia, Gaia-Sausage Enceladus, Helmi stream, Thamnos-1, and Thamnos-2 (which could be members of past Galactic merger events) are shown as shaded regions, and the members of dynamically tagged groups (DTGs) classified by \citet{Lovdal+2022} are shown as gray triangles. We also include the other 41 stars from \citetalias{sestito2019} for comparison. We have recomputed the orbital parameters for these stars in the same Galactic potential used for \rave, using distances from \citet{Bailer-Jones+2021}, positions and proper motions from Gaia DR3, and radial velocities from Gaia DR3 or \citetalias{sestito2019} (with preference to Gaia DR3 when both are available). The UMP stars are displayed according to the original classification of the orbit by those authors: Outer Halo (blue squares), Inner Halo (violet circles), and Planar (cyan diamonds).

One of the main conclusions from Figure~\ref{kin_data} is that \rave\, does not appear to be dynamically associated to any major halo substructure, and shows no similarity with the dynamical properties of the DTGs. DTG\#3 (highlighted as darker inverted triangles) is the most similar when considering energy and angular momentum; however, its members have higher eccentricities than \rave. 

Another important result is that we confirm the planar orbit classification of 6 out of 11 stars previously classified as planar by \citetalias{sestito2019} (using the criteria of $z_\mathrm{max} < 3$ kpc), including \rave. The main difference between ours and \citetalias{sestito2019} computations is the way heliocentric distances were estimated. We rely on accurate photogeometric distances from \citet{Bailer-Jones+2021}, while \citetalias{sestito2019} were limited to less accurate isochrone-based distances. This leads us to believe the five stars which show no planar orbits were originally misclassified by those authors. In their study, they already pointed out that two of them could be classified as Inner Halo (IH) taking into account their observational uncertainties. This result suggests that even though the planar orbital nature of \rave\, is not unique, it is rarer than previously reported in the literature.

We can also further classify these six known EMP/UMP stars into two groups: i) the stars 2MASS~J18082002$-$5104378 and SDSS~J102915$+$172927 are on more circular orbits ($ecc~\leq~0.15$); ii) while \rave, HE~1012$-$1540, SDSS~J014036.21$+$234458.1, and LAMOST~J125346.09$+$075343.1 all have eccentricities between $0.69-0.83$ and also share similar $L_z$ ([540, 885] kpc~km~s$^{-1}$) and $E$ ([$-178000, -158000$] km$^2$~s$^{-2}$). This similarity suggests a possible common origin for this group of stars. These stars are highlighted in Figure~\ref{kin_data} with red-contoured diamonds and their basic information is listed in Table~\ref{table_kin_group}.

\begin{deluxetable}{@{}lccccc@{}}[!ht]
\tabletypesize{\small}
\tabletypesize{\footnotesize}
\tablewidth{0pc}
\label{table_kin_group}
\tablecaption{Parameters of \rave\, and other UMP stars with similar orbital properties. All parameters were derived in this work, except for the metallicities of SDSS~J0140, HE~1012$-$1540, LAM~J1253, which were compiled by \citet{sestito2019}. Star names SDSS~J0140 and LAM~J1253 are abbreviated names for SDSS~J014036.21$+$234458.1 and LAMOST~J125346.09$+$075343.1, respectively.}
\tablehead{
\colhead{Star}&
\colhead{$ecc$}&
\colhead{$z_\mathrm{max}$}&
\colhead{$E/10^5$}&
\colhead{$L_z/10^3$} &
\colhead{[Fe/H]}\\
\colhead{}&
\colhead{}&
\colhead{kpc}&
\colhead{km$^2$ s$^{-2}$}&
\colhead{kpc~km$^{-1}$}&
\colhead{dex}}
\startdata
\protect\rave  & 0.73 & 1.28 & $-$1.78 & 0.541 & $-$3.96 $\pm$ 0.09 \\
SDSS~J0140     & 0.70 & 1.85 & $-$1.64 & 0.796 & $-$4.00 $\pm$ 0.30 \\
HE~1012$-$1540 & 0.83 & 1.74 & $-$1.58 & 0.558 & $-$4.17 $\pm$ 0.16 \\
LAM~J1253      & 0.69 & 1.27 & $-$1.61 & 0.885 & $-$4.02 $\pm$ 0.06 \\
\enddata
\end{deluxetable}

\subsection{Age and Initial Mass}

\begin{figure}[!ht]
 \includegraphics[width=\linewidth]{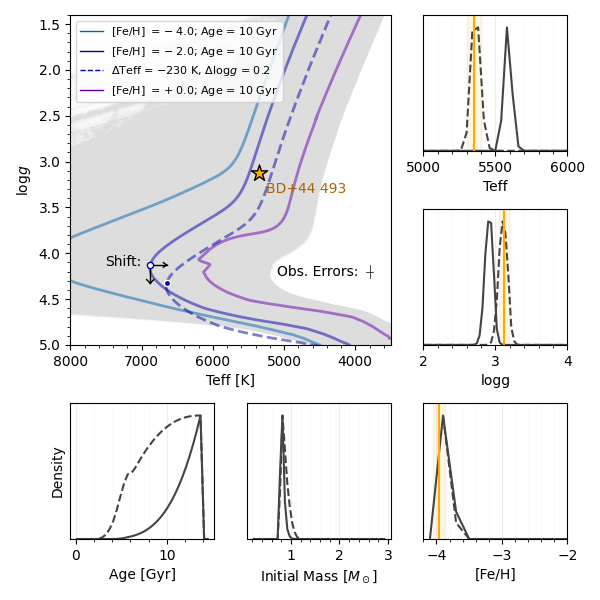}
     \caption{Top left: Kiel diagram for \rave\, (yellow star-shaped symbol) and the MIST model grid used in this work (grey area). For guidance, we highlight three isochrones of 10 Gyr and three different metallicities ($-$4.0, violet; $-$2.0, blue; 0.0, cyan). We also show the offset applied to the models (for simplicity, this is shown only for the $-2.0$ metallicity isochrone, represented by a dashed blue line). The observational errors for \rave\, are shown in the bottom-right corner. The right-side panels display the PDFs obtained by our method for effective temperature, surface gravity, and metallicity for the original models (solid lines) and the offset models (dashed lines), along with the true observational values (vertical yellow lines). The bottom-left and bottom-center panels display the obtained age and initial mass PDFs, for the original and offset models (solid and dashed lines, respectively).}
 \label{isocdata}
\end{figure}

We estimated the age and initial mass of \rave\, using the Bayesian technique described in \citet{Almeida-Fernandes+23}. The methodology is based on estimating the likelihood of observing the measured atmospheric parameters given the predictions of synthetic models of different ages, initial stellar masses, and chemical abundances, as well as on prior distributions for these parameters. The multi-dimensional Bayesian posterior probability density function (PDF) is then normalized with respect to each parameter to obtain their respective 1-dimensional PDFs.

For this work, we used models from the MESA Isochrones \& Stellar Tracks \citep[MIST][]{Dotter+2016}, which were chosen due to covering metallicities down to [M/H] = $-4$. Our isochrone grid ranges from 0.1 to 15.0 Gyr in steps of 0.2 Gyr for ages, and from $-$4.0 to $+$0.5 by 0.1~dex for metallicities. This grid covers the entire parameter space around the measured atmospheric parameters for \rave, represented by the gray area in Figure~\ref{isocdata}. 

When compared to the temperature and surface gravity of \rave, the three 10-Gyr isochrones shown in Figure~\ref{isocdata}, for metallicities $-4.0$, $-2.0$, and 0.0, indicate that the models overestimate the temperature of ultra-metal-poor stars. This is confirmed by the first run of our methodology, which results in the solid-line PDFs shown in the right and bottom panels of Figure~\ref{isocdata}. There is a clear offset between these PDFs and the observed parameters (represented by the yellow vertical line). Based on this first run, we applied an offset of $-230$~K and 0.2~dex for the temperature and surface gravity of all models in our grid and repeated the process of generating the 1-D PDFs. This correction is represented in Figure~\ref{isocdata} by showing the result of applying this offset to the 10 Gyr isochrone with [Fe/H] = $-2$ as an example. It is possible to see that the PDFs obtained using the offset-corrected models (dashed lines in the right column panels) agree with the observed atmospheric parameters (\teff, \logg, and \metal).

We characterized the age and initial mass of \rave\, from the PDFs obtained using the offset-correct models, which are shown as dashed lines in the bottom-left and bottom-center panels of Figure~\ref{isocdata}. As expected for stars in the evolutionary phase of \rave, the age-PDF has a broad profile, resulting in large errors for this parameter. The obtained PDF results in a median age of 10.3~Gyr, with 16th and 84th percentiles given by 6.8 and 13.2 Gyr, respectively. The PDF is much better constrained for the initial mass, resulting in a median value of 0.83 \Msun, and 16th and 84th percentiles of 0.78 and 0.92 \Msun.

\subsection{Kinematical Age}

We have also estimated a kinematical-based age for \rave\, from its heliocentric Galactic $U$, $V$, $W$ velocities. The dispersion of these velocities for a group of stars has long been known to correlate with stellar ages \citep{Wielen1977}. Based on this characteristic, \citet{Almeida-Fernandes+2018a, Almeida-Fernandes+2018b} developed a Bayesian method to obtain a PDF for the stellar age based on the likelihood of observing the measured velocities given a parameterized velocity ellipsoid for different age groups.

\begin{figure}[!ht]
 \includegraphics[width=\linewidth]{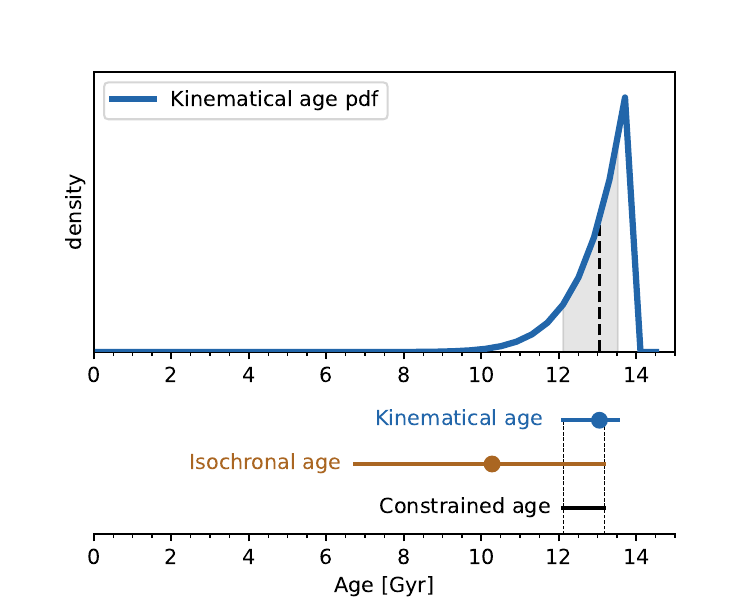}
     \caption{Top: Age PDF obtained for \rave\, from the heliocentric $UVW$ Galactic velocities using the method from \citet{Almeida-Fernandes+2018a}. Bottom: The median age obtained by the kinematical and isochronal methods is represented by blue and golden circles, respectively. The blue and golden lines represent the 16th and 84th age percentiles obtained for each method. The constrained age interval (shown as a black line) corresponds to the intersection of the age intervals from the previous methods.}
 \label{kin_age}
\end{figure}

Adopting the velocity ellipsoid parametrization of \citet{Almeida-Fernandes+2018a}, we obtained the age PDF for \rave\, shown in Figure~\ref{kin_age}. This PDF results in a median age of 13.0~Gyr, with lower and upper limits of 12.1 and 13.7~Gyr based on the 16th and 84th percentiles of the distribution.
This method does not depend on the intrinsic evolutionary stage of a given star and can be considered statistically independent from the isochronal age derived in the previous section. For this reason, we can constrain the age by taking the overlap between the upper and lower limits of each estimation. This approach is represented at the bottom of Figure~\ref{kin_age}, and results in a well-constrained age interval of [12.1, 13.2] Gyr for \rave.

\vspace{2cm}

\subsection{The chrono-chemo-dynamical nature of \protect\rave}

Our analysis indicate that \rave\, is an old ([12.1, 13.2] Gyr) EMP/UMP star (\metal $= -3.96 \pm 0.09$) residing in a planar orbit ($z_\mathrm{max} = 1.286$ kpc, $R_\mathrm{apo} = 8.623$ kpc). \citet{sestito2019} proposed three scenarios to explain the nature of such a star: i) the star was formed in a quasi-circular orbit as the \ion{H}{1} disk settled, and its orbit later evolved due to kinematic heating in the disk \citep[e.g.][]{nordstrom2004}. This scenario aligns with recent findings that suggest the Milky Way's thick disk began forming $\sim$~13~Gyr ago \citep{Xiang+Rix2022}, as possibly did the thin disk \citep{Nepal+2024}. In this case, \rave\, could belong to one of the first generations of stars to form in these structures; ii) the star was accreted from a massive satellite dwarf galaxy whose orbit was aligned with the Milky Way plane. However, our results do not indicate any orbital similarity between this star and the remnants of the known major merger events observed in the Galactic halo. Additionally, \rave\, does not appear to be associated with the Atari disk \citep{mardini2022}, as it has lower metallicity, higher eccentricity, and a higher apogalactic radius compared to stars linked with this substructure; iii) the last scenario suggests that this and other UMP stars with planar orbits may have belonged to one of the building blocks of the proto-Milky Way. 

The similarity in orbital characteristics among the other three UMP stars considered in this work (listed in Table~\ref{table_kin_group}) could indicate they originated from the same building block. As mentioned above, this common building block could be a yet unknown early dwarf galaxy that merged with the forming Milky Way. In addition to having metallicity values in the \metal$\sim-4$ regime with a 0.2~dex scatter, these four stars are all CEMP-no with $+1.10\leq$\cfe$\leq+2.20$. A detailed comparison of their chemical abundance patterns and possible progenitor populations \citep[similar to][]{placco2016b} can help further speculate on a possible common origin. Nonetheless, a much larger sample of UMP stars is needed to develop a clearer understanding of this scenario.

\section{Constraining Planetary Masses around \protect\rave}
\label{planet}

As described throughout the text, \rave\, is the brightest\footnote{The next brightest stars with \metal$\lesssim-4.0$ are CD$-$38~245 \citep{roederer2014} and 2MASS~J18082002$-$5104378 \citep{melendez2016,mardini2022}, both with $V=11.93$.} ($V=9.075$) example of a class of rare objects (UMP; \metal$\lesssim-4.0$) with less than 40 stars found to date in the Galaxy. As such, \rave\, is a prime candidate for a new avenue of exploration in stellar archaeology: planet formation at the lowest metallicities. 
As noted in the Introduction, radial velocity precision scales with the information content of the stellar spectrum.
Even though the number of absorption features available for measurement in the spectrum of \rave\, is limited when compared to higher metallicity stars with similar temperatures, it is still possible to place meaningful constraints on possible exoplanet companion masses due to the exquisite quality of the NEID data.

\begin{figure}[!ht]
 \includegraphics[width=1.1\linewidth]{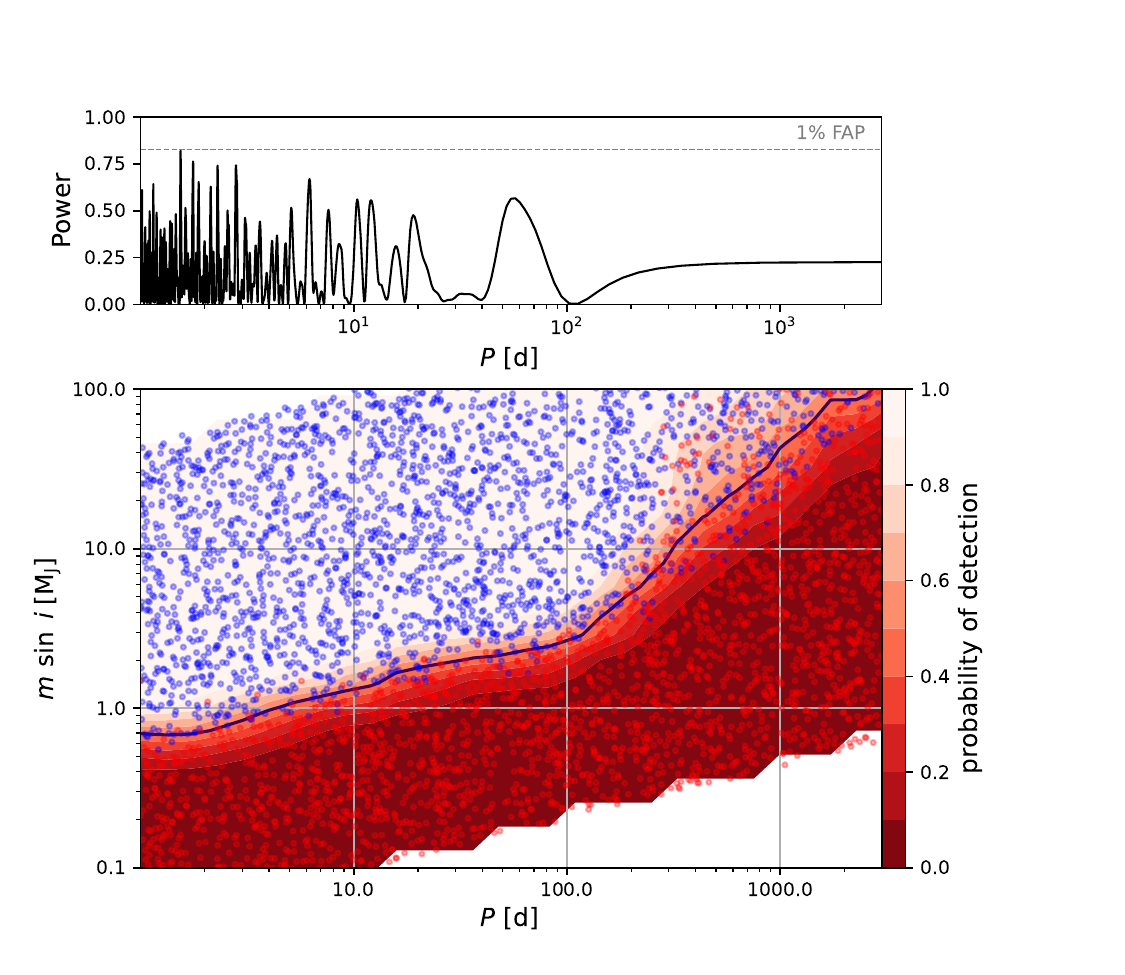}
     \caption{Upper panel: Generalized Lomb-Scargle periodogram of the NEID RV measurements of \rave. No signals are detected above the 1\% false alarm level threshold. Lower panel: Results of injection-recovery tests for NEID RV measurements of \rave. Blue points are injected planets that are successfully recovered, and red points are those that would not be detected. The background shading indicates the recovery fraction across the mass-period grid that was tested, where white corresponds to a 100\% recovery rate and deep red is a 0\% recovery rate. We also show the 50\% recovery threshold as a black line.} 
 \label{fig:completeness}
\end{figure}

We show the generalized Lomb-Scargle \citep[GLS;][]{Zechmeister2009} periodogram of the NEID RVs in Figure~\ref{fig:completeness}.
No signals are detected above a false alarm level of 1\%. To assess our sensitivity to substellar companions orbiting \rave, we use the \texttt{RVSearch} package\footnote{\url{https://california-planet-search.github.io/rvsearch/}} to perform injection-recovery tests on the NEID data. The \texttt{RVSearch} methods are described in detail in \citet{Rosenthal2021}. In brief, we inject a Keplerian signal into the NEID RV time series and calculate the Bayesian Information Criterion (BIC) for fixed-period sinusoidal fits across a range of orbital periods. \texttt{RVSearch} then computes the change in BIC, or $\Delta$BIC, between a zero-planet model and this sinusoidal approximation of a single-planet model. A periodogram is constructed from all $\Delta$BIC values on our period grid, and a more refined fit using a full Keplerian is performed for any signal that exceeds a 0.1\% false alarm probability detection threshold. The injected planet is considered to be detected if (a) a signal exceeds this threshold and (b) the period and semi-amplitude from the full Keplerian fit are within 25\% of the injected value.

We run our injection-recovery tests for 5000 trials with periods and semi-amplitudes drawn from log-uniform distributions with bounds 1 d $\leq P\leq$ 3000 d and $10$ m~s$^{-1} \leq K \leq 10$ km~s$^{-1}$, and orbital phases and eccentricities drawn from uniform distributions with bounds $0<\phi\leq2\pi$ and $0\leq e\leq 0.1$. The detection completeness in mass-period space is shown in Figure~\ref{fig:completeness}. The NEID data are sensitive to hot Jupiters with masses $>1$M$_{\rm J}$, and planets as small as $m \sin i = 2$M$_{\rm J}$ out to periods of 100 days. The sensitivity decreases as the orbital periods surpass our 117-d observing baseline, and the lower limit is firmly in the brown dwarf regime for periods longer than 1 year.

We also repeat the injection-recovery tests with looser eccentricity constraints, but we find that the search completeness drops off rapidly at more eccentric orbits, with the recovery fraction dropping below $0.5$ at $e=0.25$  for our entire mass-period grid. This is consistent with expectations for blind RV searches. The RV change in a Keplerian signal is more tightly phase-constrained for eccentric orbits relative to circular orbits, so it is less likely that one will capture significant variations without \emph{a priori} knowledge of the orbital period and phase.


\section{Conclusions and Future Work}
\label{conclusion}

In this work, we presented the chemo-dynamical analysis of \rave\, and provided for the first time constraints on planetary masses around EMP/UMP stars. The high-resolution, high-S/N NEID spectra allowed for the determination of accurate stellar atmospheric parameters and chemical abundances/upper limits for 17 elements, complemented with 11 abundances re-determined from Subaru and Hubble data. We compared the light-element abundance pattern of \rave, using both LTE and NLTE abundances, with a grid of Pop. III supernova models and confirm previous results suggesting its progenitor to be a metal-free 20.5\,M$_\odot$, star in the early Universe. Our estimates of orbital parameters, mass (0.83$^{+0.09}_{-0.05}~\rm{M}_{\odot}$) and age ([12.1, 13.2] Gyr) also corroborate the hypothesis that \rave\, is a bonafide second-generation star that is not associated with any major merging event in the early history of the Milky Way.
We present a first-of-its-kind analysis of RV companion sensitivity limits for an EMP host star. Though we do not detect any signals, we show that it is possible for RV observations to probe the planetary mass regime despite the low spectral information content of a star with \metal$\approx-4$, opening a new avenue of exploration and synergy between stellar archaeology and exoplanet science. Additional observations and a longer observing baseline for \rave\, and other bright low-metallicity stars would extend our mass and period sensitivity even further.

\begin{acknowledgments}

The authors would like to thank Wako Aoki for providing the Subaru/HDS data for \rave\, and the anonymous referee for the insightful comments on the manuscript.
The work of V.M.P., A.F.G, S.E.L., J.R., S.R., and J.D.C. is supported by NOIRLab, which is managed by the Association of Universities for Research in Astronomy (AURA) under a cooperative agreement with the U.S. National Science Foundation.
F.A.-F. acknowledges funding for this work from FAPESP grants 2018/20977-2 and
2021/09468-1.
I.U.R.\ acknowledges support
from U.S.\ National Science Foundation (NSF) grants  
PHY~14-30152 (Physics Frontier Center/JINA-CEE), 
AST~2205847, and
the NASA Astrophysics Data Analysis Program,
grant 80NSSC21K0627.
Based on observations at NSF Kitt Peak National Observatory, NSF NOIRLab (Prop.
ID 2023B-879248; PIs: J. Rajagopal and V. Placco), managed by the Association of
Universities for Research in Astronomy (AURA) under a cooperative agreement with
the U.S. National Science Foundation. The authors are honored to be permitted to
conduct astronomical research on I'oligam Du’ag (Kitt Peak), a mountain with
particular significance to the Tohono O’odham. %
This research has made use of NASA's Astrophysics Data System Bibliographic
Services; the arXiv pre-print server operated by Cornell University; the
{\texttt{SIMBAD}} database hosted by the Strasbourg Astronomical Data Center;
and the online Q\&A platform {\texttt{stackoverflow}}
(\href{http://stackoverflow.com/}{http://stackoverflow.com/}).

\end{acknowledgments}

\software{
{\texttt{astropy}}\,\citep{astropy2013,astropy2018}, 
{\texttt{awk}}\,\citep{awk}, 
{\texttt{dustmaps}}\,\citep{green2018}, 
{\texttt{gnuplot}}\,\citep{gnuplot}, 
{\texttt{NOIRLab IRAF}}\,\citep{tody1986,tody1993,fitzpatrick2024}, 
{\texttt{linemake}}\,\citep{placco2021,placco2021a},
{\texttt{matplotlib}}\,\citep{matplotlib}, 
{\texttt{MOOG}}\,\citep{sneden1973},  
{\texttt{numpy}}\,\citep{numpy}, 
{\texttt{pandas}}\,\citep{pandas}, 
{\texttt{RVSearch}}\,\citep{Rosenthal2021}, 
{\texttt{sed}}\,\citep{sed},
{\texttt{stilts}}\,\citep{stilts}.
}

\facilities{
WIYN (NEID)
}

\vfill
\newpage

\bibliographystyle{aasjournal}
\bibliography{bibliografia}

\twocolumngrid

\startlongtable

\begin{deluxetable}{lrrrrrr} 
\tabletypesize{\tiny}
\tablewidth{0pc}
\tablecaption{\label{eqwl} Atomic Data and Derived Abundances}
\tablehead{
\colhead{Ion}&
\colhead{$\lambda$}&
\colhead{$\chi$} &
\colhead{$\log\,gf$}&
\colhead{$EW$}&
\colhead{$\log\epsilon$\,(X)}&
\colhead{$\Delta$}\\
\colhead{}&
\colhead{({\AA})}&
\colhead{(eV)} &
\colhead{}&
\colhead{(m{\AA})}&
\colhead{}&
\colhead{NLTE}}
\startdata
\ion{Li}{1}  & 6707.80 & 0.00    &    0.17 &    syn &    0.86 &     0.006 \\
CH           & 4313.00 & \nodata & \nodata &    syn &    5.87 &   \nodata \\
\ion{Na}{1}  & 5889.95 & 0.00    &    0.11 &    syn &    2.63 &  $-$0.107 \\
\ion{Na}{1}  & 5895.92 & 0.00    & $-$0.19 &    syn &    2.58 &  $-$0.052 \\
\ion{Mg}{1}  & 3829.35 & 2.71    & $-$0.23 &  91.12 &    4.13 &     0.197 \\
\ion{Mg}{1}  & 3832.30 & 2.71    &    0.25 & 121.40 &    4.17 &     0.148 \\
\ion{Mg}{1}  & 3838.29 & 2.72    &    0.47 & 137.95 &    4.17 &     0.126 \\
\ion{Mg}{1}  & 4702.99 & 4.33    & $-$0.44 &  12.49 &    4.37 &     0.150 \\
\ion{Mg}{1}  & 5172.68 & 2.71    & $-$0.36 &  97.37 &    4.17 &     0.170 \\
\ion{Mg}{1}  & 5183.60 & 2.72    & $-$0.17 & 109.89 &    4.18 &     0.154 \\
\ion{Mg}{1}  & 5528.40 & 4.35    & $-$0.55 &  11.59 &    4.44 &     0.175 \\
\ion{Mg}{1}  & 8806.76 & 4.35    & $-$0.14 &  29.13 &    4.39 &     0.108 \\
\ion{Al}{1}  & 3961.52 & 0.01    & $-$0.33 &    syn &    1.94 &     0.600 \\
\ion{Si}{1}  & 3905.52 & 1.91    & $-$1.04 &    syn &    3.90 &     0.111 \\
\ion{Ca}{1}  & 4226.74 & 0.00    &    0.24 &  85.65 &    2.50 &     0.234 \\
\ion{Ca}{1}  & 4318.65 & 1.90    & $-$0.21 &   3.36 &    2.67 &     0.000 \\
\ion{Ca}{1}  & 4434.96 & 1.89    & $-$0.06 &   6.50 &    2.81 &     0.166 \\
\ion{Ca}{1}  & 4454.78 & 1.90    &    0.26 &  10.70 &    2.73 &     0.238 \\
\ion{Ca}{1}  & 6122.22 & 1.89    & $-$0.33 &   4.37 &    2.82 &   \nodata \\
\ion{Ca}{1}  & 6162.17 & 1.90    & $-$0.11 &   6.10 &    2.76 &     0.239 \\
\ion{Ca}{1}  & 6439.07 & 2.52    &    0.33 &   4.26 &    2.80 &     0.179 \\
\ion{Sc}{2}  & 4246.82 & 0.32    &    0.24 &    syn & $-$0.61 &   \nodata \\
\ion{Ti}{1}  & 3989.76 & 0.02    & $-$0.13 &   4.01 &    1.33 &   \nodata \\
\ion{Ti}{1}  & 3998.64 & 0.05    &    0.02 &   4.74 &    1.29 &   \nodata \\
\ion{Ti}{1}  & 4533.24 & 0.85    &    0.54 &   3.28 &    1.41 &   \nodata \\
\ion{Ti}{1}  & 4534.78 & 0.84    &    0.35 &   2.50 &    1.47 &   \nodata \\
\ion{Ti}{1}  & 4981.73 & 0.84    &    0.57 &   3.30 &    1.34 &   \nodata \\
\ion{Ti}{1}  & 4991.07 & 0.84    &    0.45 &   2.61 &    1.36 &   \nodata \\
\ion{Ti}{1}  & 4999.50 & 0.83    &    0.32 &   2.80 &    1.51 &   \nodata \\
\ion{Ti}{2}  & 3759.29 & 0.61    &    0.28 &  68.32 &    1.12 &     0.127 \\
\ion{Ti}{2}  & 3761.32 & 0.57    &    0.18 &  68.95 &    1.19 &     0.119 \\
\ion{Ti}{2}  & 3913.46 & 1.12    & $-$0.36 &  20.92 &    1.16 &     0.104 \\
\ion{Ti}{2}  & 4395.03 & 1.08    & $-$0.54 &  20.18 &    1.22 &     0.063 \\
\ion{Ti}{2}  & 4417.71 & 1.17    & $-$1.19 &   4.30 &    1.21 &     0.070 \\
\ion{Ti}{2}  & 4443.80 & 1.08    & $-$0.71 &  15.08 &    1.23 &     0.007 \\
\ion{Ti}{2}  & 4450.48 & 1.08    & $-$1.52 &   2.83 &    1.25 &     0.046 \\
\ion{Ti}{2}  & 4501.27 & 1.12    & $-$0.77 &  12.17 &    1.22 &     0.115 \\
\ion{Ti}{2}  & 4533.96 & 1.24    & $-$0.53 &  14.14 &    1.18 &     0.096 \\
\ion{Ti}{2}  & 4571.97 & 1.57    & $-$0.31 &  10.37 &    1.16 &     0.025 \\
\ion{Cr}{1}  & 4274.80 & 0.00    & $-$0.22 &  16.61 &    1.24 &     0.815 \\
\ion{Cr}{1}  & 4289.72 & 0.00    & $-$0.37 &  14.31 &    1.31 &     0.815 \\
\ion{Mn}{1}  & 4030.75 & 0.00    & $-$0.50 &   6.65 &    0.43 &     0.956 \\
\ion{Mn}{1}  & 4033.06 & 0.00    & $-$0.65 &   4.23 &    0.37 &     0.950 \\
\ion{Fe}{1}  & 3743.36 & 0.99    & $-$0.78 &  59.65 &    3.55 &     0.221 \\
\ion{Fe}{1}  & 3758.23 & 0.96    & $-$0.01 &  85.25 &    3.46 &     0.187 \\
\ion{Fe}{1}  & 3763.79 & 0.99    & $-$0.22 &  75.62 &    3.43 &     0.213 \\
\ion{Fe}{1}  & 3765.54 & 3.24    &    0.48 &  12.22 &    3.47 &     0.191 \\
\ion{Fe}{1}  & 3767.19 & 1.01    & $-$0.39 &  69.35 &    3.44 &     0.221 \\
\ion{Fe}{1}  & 3786.68 & 1.01    & $-$2.18 &   9.43 &    3.63 &     0.197 \\
\ion{Fe}{1}  & 3787.88 & 1.01    & $-$0.84 &  56.86 &    3.55 &     0.222 \\
\ion{Fe}{1}  & 3790.09 & 0.99    & $-$1.74 &  18.68 &    3.53 &     0.201 \\
\ion{Fe}{1}  & 3805.34 & 3.30    &    0.31 &   7.51 &    3.46 &     0.194 \\
\ion{Fe}{1}  & 3808.73 & 2.56    & $-$1.17 &   2.66 &    3.69 &     0.207 \\
\ion{Fe}{1}  & 3812.96 & 0.96    & $-$1.03 &  49.39 &    3.50 &     0.207 \\
\ion{Fe}{1}  & 3815.84 & 1.48    &    0.24 &  74.16 &    3.42 &     0.219 \\
\ion{Fe}{1}  & 3820.43 & 0.86    &    0.16 &  98.73 &    3.48 &     0.165 \\
\ion{Fe}{1}  & 3824.44 & 0.00    & $-$1.36 &  78.94 &    3.62 &     0.205 \\
\ion{Fe}{1}  & 3825.88 & 0.91    & $-$0.02 &  88.06 &    3.45 &     0.178 \\
\ion{Fe}{1}  & 3827.82 & 1.56    &    0.09 &  64.22 &    3.38 &     0.222 \\
\ion{Fe}{1}  & 3840.44 & 0.99    & $-$0.50 &  68.13 &    3.47 &     0.220 \\
\ion{Fe}{1}  & 3841.05 & 1.61    & $-$0.04 &  56.78 &    3.37 &     0.220 \\
\ion{Fe}{1}  & 3849.97 & 1.01    & $-$0.86 &  56.51 &    3.54 &     0.158 \\
\ion{Fe}{1}  & 3850.82 & 0.99    & $-$1.74 &  21.58 &    3.60 &     0.158 \\
\ion{Fe}{1}  & 3856.37 & 0.05    & $-$1.28 &  82.50 &    3.68 &     0.204 \\
\ion{Fe}{1}  & 3859.91 & 0.00    & $-$0.71 & 104.88 &    3.64 &     0.167 \\
\ion{Fe}{1}  & 3865.52 & 1.01    & $-$0.95 &  52.23 &    3.53 &     0.216 \\
\ion{Fe}{1}  & 3878.02 & 0.96    & $-$0.90 &  57.99 &    3.56 &     0.208 \\
\ion{Fe}{1}  & 3878.57 & 0.09    & $-$1.36 &  78.27 &    3.67 &     0.208 \\
\ion{Fe}{1}  & 3895.66 & 0.11    & $-$1.67 &  66.93 &    3.66 &     0.218 \\
\ion{Fe}{1}  & 3899.71 & 0.09    & $-$1.52 &  73.18 &    3.67 &     0.214 \\
\ion{Fe}{1}  & 3902.95 & 1.56    & $-$0.44 &  47.43 &    3.49 &     0.213 \\
\ion{Fe}{1}  & 3906.48 & 0.11    & $-$2.20 &  44.60 &    3.63 &     0.216 \\
\ion{Fe}{1}  & 3917.18 & 0.99    & $-$2.15 &   9.65 &    3.58 &     0.171 \\
\ion{Fe}{1}  & 3920.26 & 0.12    & $-$1.73 &  64.84 &    3.67 &     0.219 \\
\ion{Fe}{1}  & 3922.91 & 0.05    & $-$1.63 &  71.44 &    3.68 &     0.219 \\
\ion{Fe}{1}  & 3927.92 & 0.11    & $-$1.52 &  72.84 &    3.68 &     0.215 \\
\ion{Fe}{1}  & 3930.30 & 0.09    & $-$1.49 &  74.54 &    3.67 &     0.215 \\
\ion{Fe}{1}  & 3949.95 & 2.18    & $-$1.25 &   3.68 &    3.50 &     0.212 \\
\ion{Fe}{1}  & 3956.68 & 2.69    & $-$0.43 &   7.77 &    3.56 &     0.196 \\
\ion{Fe}{1}  & 3977.74 & 2.20    & $-$1.12 &   4.82 &    3.51 &     0.214 \\
\ion{Fe}{1}  & 4005.24 & 1.56    & $-$0.58 &  41.66 &    3.49 &     0.207 \\
\ion{Fe}{1}  & 4045.81 & 1.49    &    0.28 &  78.11 &    3.44 &     0.206 \\
\ion{Fe}{1}  & 4063.59 & 1.56    &    0.06 &  67.33 &    3.44 &     0.217 \\
\ion{Fe}{1}  & 4071.74 & 1.61    & $-$0.01 &  61.71 &    3.42 &     0.217 \\
\ion{Fe}{1}  & 4118.55 & 3.57    &    0.21 &   5.15 &    3.63 &     0.194 \\
\ion{Fe}{1}  & 4132.06 & 1.61    & $-$0.68 &  36.77 &    3.53 &     0.208 \\
\ion{Fe}{1}  & 4134.68 & 2.83    & $-$0.65 &   3.14 &    3.49 &     0.226 \\
\ion{Fe}{1}  & 4143.41 & 3.05    & $-$0.20 &   4.19 &    3.41 &     0.210 \\
\ion{Fe}{1}  & 4143.87 & 1.56    & $-$0.51 &  45.28 &    3.48 &     0.210 \\
\ion{Fe}{1}  & 4147.67 & 1.49    & $-$2.07 &   4.23 &    3.62 &     0.003 \\
\ion{Fe}{1}  & 4181.76 & 2.83    & $-$0.37 &   6.07 &    3.51 &     0.230 \\
\ion{Fe}{1}  & 4187.04 & 2.45    & $-$0.56 &  10.83 &    3.58 &     0.192 \\
\ion{Fe}{1}  & 4187.80 & 2.42    & $-$0.51 &  10.24 &    3.47 &     0.192 \\
\ion{Fe}{1}  & 4191.43 & 2.47    & $-$0.67 &   6.74 &    3.48 &     0.192 \\
\ion{Fe}{1}  & 4199.10 & 3.05    &    0.16 &  13.42 &    3.60 &     0.197 \\
\ion{Fe}{1}  & 4202.03 & 1.49    & $-$0.69 &  41.79 &    3.51 &     0.209 \\
\ion{Fe}{1}  & 4216.18 & 0.00    & $-$3.36 &   8.21 &    3.61 &     0.217 \\
\ion{Fe}{1}  & 4227.43 & 3.33    &    0.27 &   7.10 &    3.47 &     0.184 \\
\ion{Fe}{1}  & 4233.60 & 2.48    & $-$0.60 &   7.82 &    3.49 &     0.192 \\
\ion{Fe}{1}  & 4250.12 & 2.47    & $-$0.38 &  12.46 &    3.48 &     0.204 \\
\ion{Fe}{1}  & 4250.79 & 1.56    & $-$0.71 &  37.07 &    3.50 &     0.204 \\
\ion{Fe}{1}  & 4260.47 & 2.40    &    0.08 &  30.12 &    3.45 &     0.199 \\
\ion{Fe}{1}  & 4271.15 & 2.45    & $-$0.34 &  17.35 &    3.59 &     0.210 \\
\ion{Fe}{1}  & 4271.76 & 1.49    & $-$0.17 &  65.26 &    3.51 &     0.210 \\
\ion{Fe}{1}  & 4282.40 & 2.18    & $-$0.78 &   9.46 &    3.44 &     0.207 \\
\ion{Fe}{1}  & 4325.76 & 1.61    &    0.01 &  64.90 &    3.44 &     0.217 \\
\ion{Fe}{1}  & 4375.93 & 0.00    & $-$3.00 &  19.09 &    3.66 &     0.218 \\
\ion{Fe}{1}  & 4383.55 & 1.49    &    0.21 &  79.07 &    3.46 &     0.223 \\
\ion{Fe}{1}  & 4404.75 & 1.56    & $-$0.15 &  62.98 &    3.49 &     0.236 \\
\ion{Fe}{1}  & 4415.12 & 1.61    & $-$0.62 &  41.43 &    3.54 &     0.221 \\
\ion{Fe}{1}  & 4427.31 & 0.05    & $-$2.92 &  19.43 &    3.64 &     0.219 \\
\ion{Fe}{1}  & 4442.34 & 2.20    & $-$1.23 &   4.59 &    3.56 &     0.232 \\
\ion{Fe}{1}  & 4447.72 & 2.22    & $-$1.36 &   3.51 &    3.58 &     0.234 \\
\ion{Fe}{1}  & 4459.12 & 2.18    & $-$1.31 &   4.13 &    3.56 &     0.238 \\
\ion{Fe}{1}  & 4461.65 & 0.09    & $-$3.19 &  11.98 &    3.70 &     0.220 \\
\ion{Fe}{1}  & 4466.55 & 2.83    & $-$0.60 &   3.74 &    3.50 &     0.229 \\
\ion{Fe}{1}  & 4476.02 & 2.85    & $-$0.82 &   2.26 &    3.51 &     0.216 \\
\ion{Fe}{1}  & 4494.56 & 2.20    & $-$1.14 &   5.92 &    3.58 &     0.239 \\
\ion{Fe}{1}  & 4528.61 & 2.18    & $-$0.85 &  10.78 &    3.55 &     0.242 \\
\ion{Fe}{1}  & 4531.15 & 1.48    & $-$2.10 &   3.10 &    3.47 &     0.227 \\
\ion{Fe}{1}  & 4602.94 & 1.49    & $-$2.21 &   2.96 &    3.56 &     0.016 \\
\ion{Fe}{1}  & 4871.32 & 2.87    & $-$0.34 &   5.13 &    3.40 &     0.199 \\
\ion{Fe}{1}  & 4872.14 & 2.88    & $-$0.57 &   2.86 &    3.38 &     0.199 \\
\ion{Fe}{1}  & 4890.76 & 2.88    & $-$0.38 &   5.16 &    3.45 &     0.200 \\
\ion{Fe}{1}  & 4891.49 & 2.85    & $-$0.11 &   9.63 &    3.44 &     0.200 \\
\ion{Fe}{1}  & 4918.99 & 2.86    & $-$0.34 &   5.90 &    3.45 &     0.200 \\
\ion{Fe}{1}  & 4920.50 & 2.83    &    0.07 &  14.46 &    3.44 &     0.203 \\
\ion{Fe}{1}  & 4957.30 & 2.85    & $-$0.41 &   5.27 &    3.46 &     0.206 \\
\ion{Fe}{1}  & 4957.60 & 2.81    &    0.23 &  20.40 &    3.44 &     0.206 \\
\ion{Fe}{1}  & 4994.13 & 0.92    & $-$2.97 &   2.37 &    3.58 &   \nodata \\
\ion{Fe}{1}  & 5006.12 & 2.83    & $-$0.62 &   3.61 &    3.47 &     0.200 \\
\ion{Fe}{1}  & 5012.07 & 0.86    & $-$2.64 &   6.49 &    3.65 &     0.225 \\
\ion{Fe}{1}  & 5041.76 & 1.49    & $-$2.20 &   3.04 &    3.54 &     0.072 \\
\ion{Fe}{1}  & 5049.82 & 2.28    & $-$1.36 &   2.93 &    3.53 &     0.228 \\
\ion{Fe}{1}  & 5051.63 & 0.92    & $-$2.76 &   3.84 &    3.58 &     0.229 \\
\ion{Fe}{1}  & 5083.34 & 0.96    & $-$2.84 &   2.68 &    3.54 &     0.229 \\
\ion{Fe}{1}  & 5150.84 & 0.99    & $-$3.04 &   1.48 &    3.51 &   \nodata \\
\ion{Fe}{1}  & 5166.28 & 0.00    & $-$4.12 &   2.00 &    3.64 &   \nodata \\
\ion{Fe}{1}  & 5171.60 & 1.49    & $-$1.72 &   8.43 &    3.52 &     0.222 \\
\ion{Fe}{1}  & 5191.45 & 3.04    & $-$0.55 &   2.85 &    3.51 &     0.209 \\
\ion{Fe}{1}  & 5192.34 & 3.00    & $-$0.42 &   3.58 &    3.44 &     0.209 \\
\ion{Fe}{1}  & 5194.94 & 1.56    & $-$2.02 &   4.44 &    3.59 &     0.211 \\
\ion{Fe}{1}  & 5216.27 & 1.61    & $-$2.08 &   3.42 &    3.59 &     0.220 \\
\ion{Fe}{1}  & 5232.94 & 2.94    & $-$0.06 &   8.90 &    3.43 &     0.213 \\
\ion{Fe}{1}  & 5266.56 & 3.00    & $-$0.38 &   3.58 &    3.40 &     0.211 \\
\ion{Fe}{1}  & 5269.54 & 0.86    & $-$1.32 &  58.32 &    3.71 &     0.232 \\
\ion{Fe}{1}  & 5270.36 & 1.61    & $-$1.51 &  13.61 &    3.67 &     0.232 \\
\ion{Fe}{1}  & 5324.18 & 3.21    & $-$0.11 &   4.17 &    3.41 &     0.208 \\
\ion{Fe}{1}  & 5328.04 & 0.92    & $-$1.47 &  46.98 &    3.68 &     0.192 \\
\ion{Fe}{1}  & 5328.53 & 1.56    & $-$1.85 &   6.38 &    3.58 &     0.192 \\
\ion{Fe}{1}  & 5341.02 & 1.61    & $-$1.95 &   5.12 &    3.63 &     0.000 \\
\ion{Fe}{1}  & 5371.49 & 0.96    & $-$1.64 &  34.63 &    3.64 &     0.232 \\
\ion{Fe}{1}  & 5383.37 & 4.31    &    0.64 &   2.36 &    3.56 &     0.188 \\
\ion{Fe}{1}  & 5397.13 & 0.92    & $-$1.98 &  21.65 &    3.64 &     0.234 \\
\ion{Fe}{1}  & 5405.77 & 0.99    & $-$1.85 &  23.25 &    3.62 &     0.232 \\
\ion{Fe}{1}  & 5415.20 & 4.39    &    0.64 &   1.97 &    3.57 &     0.199 \\
\ion{Fe}{1}  & 5429.70 & 0.96    & $-$1.88 &  24.49 &    3.65 &     0.232 \\
\ion{Fe}{1}  & 5434.52 & 1.01    & $-$2.13 &  13.61 &    3.63 &     0.234 \\
\ion{Fe}{1}  & 5446.92 & 0.99    & $-$1.93 &  21.76 &    3.66 &     0.229 \\
\ion{Fe}{1}  & 5455.61 & 1.01    & $-$2.09 &  14.69 &    3.63 &     0.014 \\
\ion{Fe}{1}  & 5497.52 & 1.01    & $-$2.82 &   2.57 &    3.53 &     0.000 \\
\ion{Fe}{1}  & 5501.47 & 0.96    & $-$3.05 &   2.26 &    3.65 &     0.000 \\
\ion{Fe}{1}  & 5506.78 & 0.99    & $-$2.79 &   3.50 &    3.62 &     0.004 \\
\ion{Fe}{1}  & 5572.84 & 3.40    & $-$0.31 &   2.22 &    3.52 &     0.209 \\
\ion{Fe}{1}  & 5586.76 & 3.37    & $-$0.11 &   2.82 &    3.40 &     0.212 \\
\ion{Fe}{1}  & 5615.64 & 3.33    & $-$0.14 &   4.33 &    3.58 &     0.214 \\
\ion{Fe}{2}  & 4173.45 & 2.58    & $-$2.38 &   2.27 &    3.56 &     0.000 \\
\ion{Fe}{2}  & 4233.16 & 2.58    & $-$2.02 &   6.47 &    3.67 &     0.004 \\
\ion{Fe}{2}  & 4583.83 & 2.81    & $-$1.94 &   4.55 &    3.65 &     0.005 \\
\ion{Fe}{2}  & 4923.93 & 2.89    & $-$1.21 &  10.71 &    3.38 &     0.010 \\
\ion{Fe}{2}  & 5018.43 & 2.89    & $-$1.23 &  14.49 &    3.56 &     0.000 \\
\ion{Fe}{2}  & 5169.03 & 2.89    & $-$1.14 &  18.03 &    3.58 &     0.016 \\
\ion{Co}{1}  & 3845.47 & 0.92    &    0.06 &  14.92 &    1.38 &   \nodata \\
\ion{Co}{1}  & 3995.31 & 0.92    & $-$0.18 &  10.29 &    1.42 &   \nodata \\
\ion{Co}{1}  & 4118.77 & 1.05    & $-$0.48 &   5.36 &    1.53 &   \nodata \\
\ion{Co}{1}  & 4121.32 & 0.92    & $-$0.33 &   9.92 &    1.53 &   \nodata \\
\ion{Ni}{1}  & 3783.53 & 0.42    & $-$1.40 &    syn &    2.38 &   \nodata \\
\ion{Ni}{1}  & 3807.14 & 0.42    & $-$1.23 &    syn &    2.29 &   \nodata \\
\ion{Sr}{2}  & 4077.71 & 0.00    &    0.15 &    syn & $-$1.42 &   \nodata \\
\ion{Sr}{2}  & 4215.52 & 0.00    & $-$0.17 &    syn & $-$1.40 &   \nodata \\
\ion{Ba}{2}  & 4554.03 & 0.00    &    0.16 &    syn & $-$2.53 &   \nodata \\
\ion{Ba}{2}  & 4934.10 & 0.00    & $-$0.15 &    syn & $-$2.52 &   \nodata \\
\ion{Eu}{2}  & 4205.04 & 0.00    &    0.21 &    syn &$<-$2.62 &   \nodata \\
\enddata
\tablerefs{
NLTE corrections -- 
\ion{Li}{1}: \citet{lind2009};
\ion{Na}{1}: \citet{lind2011};
\ion{Mg}{1}: \citet{bergemann2015};
\ion{Al}{1}: \citet{nordlander2017b};
\ion{Si}{1}: \citet{bergemann2013};
\ion{Ca}{1}: \citet{mashonkina2007};
\ion{Ti}{1}: \citet{bergemann2011};
\ion{Ti}{2}: \citet{bergemann2011};
\ion{Cr}{1}: \citet{bergemann2010};
\ion{Mn}{1}: \citet{bergemann2019};
\ion{Fe}{1}: \citet{bergemann2012b};
\ion{Co}{1}: \citet{bergemann2010b}
}
\end{deluxetable}


\begin{deluxetable*}{cccccr}[!ht] 
\tabletypesize{\tiny}
\tablewidth{0pc}
\tablecaption{\label{rvlit} Literature radial velocity information for \protect{\rave}}
\tablehead{
\colhead{Date}&
\colhead{Time}&
\colhead{Julian date}&
\colhead{RV}&
\colhead{$\sigma_{\rm RV}$}&
\colhead{Reference}\\
\colhead{}&
\colhead{(UTC)}&
\colhead{}&
\colhead{(km s$^{-1}$)}&
\colhead{(km s$^{-1}$)}&
\colhead{}}
\startdata
1984-01-27 & 04:54:07 & 2445726.70425 & $-$150.660 & 0.550 & \citet{carney2003} \\
1984-09-22 & 05:49:15 & 2445965.74253 & $-$151.160 & 0.770 & \citet{carney2003} \\
1985-08-24 & 08:23:20 & 2446301.84954 & $-$149.550 & 0.680 & \citet{carney2003} \\
1985-09-29 & 08:55:54 & 2446337.87215 & $-$151.570 & 0.570 & \citet{carney2003} \\
1985-10-27 & 05:13:01 & 2446365.71737 & $-$150.930 & 0.670 & \citet{carney2003} \\
1985-12-28 & 03:55:44 & 2446427.66370 & $-$150.660 & 0.500 & \citet{carney2003} \\
1986-01-19 & 03:29:26 & 2446449.64544 & $-$150.580 & 0.740 & \citet{carney2003} \\
1986-03-21 & 02:39:57 & 2446510.61108 & $-$150.670 & 0.570 & \citet{carney2003} \\
1986-10-20 & 06:50:43 & 2446723.78522 & $-$150.130 & 0.460 & \citet{carney2003} \\
1987-02-06 & 02:27:32 & 2446832.60245 & $-$150.350 & 0.550 & \citet{carney2003} \\
1987-03-11 & 02:56:22 & 2446865.62248 & $-$149.820 & 0.490 & \citet{carney2003} \\
1987-06-08 & 11:12:31 & 2446954.96702 & $-$150.660 & 0.550 & \citet{carney2003} \\
1987-06-13 & 11:30:24 & 2446959.97945 & $-$150.960 & 0.560 & \citet{carney2003} \\
1987-07-10 & 11:47:17 & 2446986.99117 & $-$149.730 & 0.480 & \citet{carney2003} \\
1987-10-08 & 08:43:32 & 2447076.86357 & $-$151.330 & 0.520 & \citet{carney2003} \\
1987-11-10 & 07:40:21 & 2447109.81969 & $-$150.510 & 0.580 & \citet{carney2003} \\
1987-12-07 & 05:46:36 & 2447136.74070 & $-$151.150 & 0.550 & \citet{carney2003} \\
1988-09-22 & 10:15:26 & 2447426.92739 & $-$150.900 & 0.440 & \citet{carney2003} \\
1988-09-29 & 09:39:57 & 2447433.90274 & $-$150.000 & 0.520 & \citet{carney2003} \\
1989-11-05 & 00:23:14 & 2447835.51613 & $-$152.410 & 0.720 & \citet{carney2003} \\
1990-01-18 & 03:40:36 & 2447909.65319 & $-$150.880 & 0.660 & \citet{carney2003} \\
1990-02-08 & 01:44:21 & 2447930.57246 & $-$149.780 & 0.600 & \citet{carney2003} \\
1992-08-23 & 05:48:16 & 2448857.74185 & $-$150.960 & 0.660 & \citet{carney2003} \\
1993-09-04 & 10:50:27 & 2449234.95170 & $-$151.330 & 0.430 & \citet{carney2003} \\
1994-09-18 & 08:04:45 & 2449613.83663 & $-$151.320 & 0.590 & \citet{carney2003} \\
1994-12-15 & 03:05:54 & 2449701.62910 & $-$152.240 & 0.520 & \citet{carney2003} \\
1997-07-30 & 07:11:03 & 2450659.79934 & $-$149.220 & 0.690 & \citet{carney2003} \\
1997-09-17 & 07:03:31 & 2450708.79411 & $-$151.070 & 0.510 & \citet{carney2003} \\
2008-08-22 & 15:26:06 & 2454701.14313 & $-$150.380 & 0.500 & \citet{ito2013} \\
2008-10-04 & 09:14:43 & 2454743.88522 & $-$149.910 & 0.500 & \citet{ito2013} \\
2008-10-05 & 11:14:35 & 2454744.96846 & $-$150.410 & 0.500 & \citet{ito2013} \\
2008-11-04 & 09:07:12 & 2454774.88000 & $-$150.000 & 0.700 & \citet{roederer2014} \\
2008-11-16 & 05:46:53 & 2454786.74089 & $-$149.940 & 0.500 & \citet{ito2013} \\
2009-05-11 & 11:02:24 & 2454962.96000 & $-$149.700 & 0.700 & \citet{roederer2014} \\
2011-08-23 & 05:31:12 & 2455796.73000 & $-$150.191 & 0.019 & \citet{hansen2016} \\
2011-09-17 & 02:24:00 & 2455821.60000 & $-$150.061 & 0.016 & \citet{hansen2016} \\
2011-10-25 & 00:57:36 & 2455859.54000 & $-$150.050 & 0.017 & \citet{hansen2016} \\
2011-11-16 & 21:21:36 & 2455882.39000 & $-$150.150 & 0.113 & \citet{hansen2016} \\
2011-11-26 & 21:50:24 & 2455892.41000 & $-$150.114 & 0.125 & \citet{hansen2016} \\
2011-12-07 & 23:45:36 & 2455903.49000 & $-$150.080 & 0.032 & \citet{hansen2016} \\
2011-12-20 & 00:00:00 & 2455915.50000 & $-$149.985 & 0.141 & \citet{hansen2016} \\
2012-02-13 & 21:21:36 & 2455971.39000 & $-$150.001 & 0.041 & \citet{hansen2016} \\
2012-08-24 & 04:48:00 & 2456163.70000 & $-$150.101 & 0.164 & \citet{hansen2016} \\
2012-09-21 & 02:38:24 & 2456191.61000 & $-$150.126 & 0.018 & \citet{hansen2016} \\
2013-01-14 & 22:19:12 & 2456307.43000 & $-$150.041 & 0.028 & \citet{hansen2016} \\
2013-02-07 & 00:00:00 & 2456330.50000 & $-$150.260 & 0.370 & \citet{starkenburg2014} \\
2013-02-16 & 00:00:00 & 2456339.50000 & $-$150.080 & 0.370 & \citet{starkenburg2014} \\
2013-07-13 & 12:00:00 & 2456487.00000 & $-$150.170 & 0.300 & \citet{starkenburg2014} \\
2013-08-04 & 12:00:00 & 2456509.00000 & $-$149.940 & 0.320 & \citet{starkenburg2014} \\
2013-08-14 & 04:33:36 & 2456518.69000 & $-$150.104 & 0.031 & \citet{hansen2016} \\
2013-08-21 & 01:54:38 & 2456525.57961 & $-$149.769 & 0.552 & \citet{arentsen2019} \\
2013-08-24 & 04:48:00 & 2456528.70000 & $-$150.081 & 0.028 & \citet{hansen2016} \\
2013-09-14 & 01:15:33 & 2456549.55246 & $-$150.237 & 0.505 & \citet{arentsen2019} \\
2013-11-07 & 03:21:36 & 2456603.64000 & $-$150.066 & 0.025 & \citet{hansen2016} \\
2014-01-27 & 22:04:48 & 2456685.42000 & $-$150.040 & 0.060 & \citet{hansen2016} \\
2014-02-08 & 17:54:48 & 2456697.24639 & $-$149.946 & 0.349 & \citet{arentsen2019} \\
2014-02-20 & 16:55:53 & 2456709.20547 & $-$149.322 & 1.322 & \citet{arentsen2019} \\
2014-02-20 & 16:57:47 & 2456709.20679 & $-$149.619 & 1.128 & \citet{arentsen2019} \\
2014-08-24 & 02:24:00 & 2456893.60000 & $-$150.127 & 0.032 & \citet{hansen2016} \\
2014-11-25 & 23:45:36 & 2456987.49000 & $-$150.101 & 0.086 & \citet{hansen2016} \\
2015-03-12 & 20:24:00 & 2457094.35000 & $-$150.100 & 0.029 & \citet{hansen2016}
\enddata
\end{deluxetable*}

\end{document}